\documentclass[structabstract]{aa}  
\usepackage{graphicx}
\usepackage{txfonts}
\usepackage{array}
\begin{document}
   \title{The VIMOS VLT Deep Survey\thanks{Based on data obtained with the European Southern Observatory Very Large
Telescope, Paranal, Chile, under Large Programs 070.A-9007 and 177.A-0837. 
Based on observations obtained with MegaPrime/MegaCam, a joint project of CFHT and CEA/DAPNIA, at the Canada-France-Hawaii Telescope (CFHT) which is operated by the National Research Council (NRC) of Canada, the Institut National des Sciences de l'Univers of the Centre National de la Recherche Scientifique (CNRS) of France, and the University of Hawaii. This work is based in part on data products produced at TERAPIX and the Canadian Astronomy Data Centre as part of the Canada-France-Hawaii Telescope Legacy Survey, a collaborative project of NRC and CNRS.}: 
the redshift distribution N(z) of magnitude-limited samples down to $i_{AB}=24.75$ and $Ks_{AB}=22$  
}

%   \subtitle{I. Overviewing the $\kappa$-mechanism}

   \author{O. Le F\`evre
          \inst{1}
          \and
P. Cassata\inst{1}
\and
O. Cucciati\inst{2}
\and
S. de la Torre\inst{3,1}
\and
B. Garilli\inst{4}
\and
O. Ilbert\inst{1}
\and
V. Le Brun\inst{1}
\and
D. Maccagni\inst{4}
\and
L. Tresse\inst{1}
\and
G. Zamorani\inst{2}
\and
S. Bardelli\inst{2}
\and
M. Bolzonella\inst{2}
\and
T. Contini\inst{5, 6}
\and
A. Iovino\inst{7}
\and
C. L\'opez-Sanjuan \inst{1,8}
\and
H.J. McCracken\inst{9}
\and
A. Pollo\inst{10, 11}
\and
L. Pozzetti\inst{2}
\and
M. Scodeggio\inst{4}
\and
L. Tasca \inst{1}
\and
D. Vergani\inst{2}
\and
A. Zanichelli\inst{12}
\and
E. Zucca\inst{2}
%\and
%et al.\inst{1}
%          C. Ptolemy\inst{2}\fnmsep\thanks{Just to show the usage
%          of the elements in the author field}
}

\institute{Aix-Marseille Universit\'e, CNRS, LAM-Laboratoire d'Astrophysique de Marseille, 38, rue Fr\'ed\'eric Joliot-Curie, 13388 Marseille, France
%              \email{olivier.lefevre@oamp.fr}
\and
INAF - Osservatorio Astronomico di Bologna, via Ranzani 1, I-
40127, Bologna, Italy
\and 
SUPA, Institute for Astronomy, University of Edinburgh, Royal Ob-
servatory, Blackford Hill, Edinburgh EH9 3HJ, UK
\and 
INAF - Istituto di Astrofisica Spaziale e Fisica Cosmica Milano, via
Bassini 15, 20133 Milano, Italy
\and
Institut de Recherche en Astrophysique et Plan\'etologie (IRAP), CNRS, 14 avenue E. Belin, 31400 Toulouse, France
\and
IRAP, Université de Toulouse, UPS-OMP, Toulouse, France
\and
INAF - Osservatorio Astronomico di Brera, Via Brera 28, 20122
Milano, via E. Bianchi 46, 23807 Merate, Italy
\and
Centro de Estudios de F\'isica del Cosmos de Arag\'on, Plaza San Juan 1, planta 2, 44001 Teruel, Spain
\and
Institut d'Astrophysique de Paris, UMR7095 CNRS,
Universit\'e Pierre et Marie Curie, 98 bis Boulevard Arago, 75014
Paris
\and
Astronomical Observatory of the Jagiellonian University, Orla 171,
30-001 Cracow, Poland
\and
National Centre for Nuclear Research, ul. Hoza 69, 00-681
Warszawa, Poland
\and
INAF - Istituto di Radioastronomia, via Gobetti 101, I-40129,
Bologna, Italy
\\ \\
             \email{olivier.lefevre@oamp.fr}
%             \thanks{The university of heaven temporarily does not
%                     accept e-mails}
             }

   \date{Received...; accepted...}

% \abstract{}{}{}{}{} 
% 5 {} token are mandatory
 
  \abstract
   {The accurate census of galaxies at different epochs since the first galaxies were formed
    is necessary to make progress in understanding galaxy evolution, including how 
    mass assembly and  star formation  evolve.}
   {We aim to measure and analyse the redshift distribution N(z) of magnitude-selected samples 
    using spectroscopic redshift measurement up to
    $z\simeq5$. }
   {We use the VIMOS VLT Deep Survey (VVDS) final data release on the 0226-04 field,
    including 10\,765 galaxies with spectroscopic redshifts, selected solely on their magnitude $17 \leq i_{AB} \leq 24.75$,
    successfully crossing any 'redshift desert'.
    We compute the redshift distribution N(z) and provide reference parametric fits  for $i-$band as  well
    as for $J$, $H$ and $Ks-$band magnitude limited samples. 
    The observed galaxy number counts in different redshift domains are compared to
    other surveys from the literature, as well as
    to results from semi-analytic models on the Millennium dark matter simulations.
}
   {The  redshift distribution of a sample with $i_{AB}\leq24$ and spectroscopic redshifts
    has a mean redshift $\bar{z}=0.92$, with 8.2\% of the galaxies 
    with $z>2$. Down to $i_{AB} \leq 24.75$ the spectroscopic redshift sample  
    has a mean redshift $\bar{z}=1.15$ and 17.1\% of the galaxies are beyond $z=2$. 
    We find that the projected sky density is $2.07\pm0.12$ galaxies
    per arcmin$^2$ at $1.4 \leq z \leq 2.5$ and $Ks_{AB}\leq22.5$, $1.72\pm0.15$ galaxies per
    arcmin$^2$ at $2.7 \leq z \leq 3.4$ and $0.59\pm0.09$ galaxies per arcmin$^2$
    at $3.4 \leq z \leq 4.5$ brighter than $i_{AB}=24.75$ (errors are including Poisson noise and cosmic variance). 
    Galaxies at $z\sim3$ identified from magnitude-selected samples are 1.5 to 3 times more numerous
    than when they are colour-colour selected, consistent with the different selection functions.
    We demonstrate that colour-colour selected samples over $1.4 \leq z \leq 4.5$
    are strongly contaminated by galaxies at other redshifts. 
    Semi-analytic models on the Millennium simulations %with WMAP1 or WMAP3 
    are found to under-predict the number of 
    luminous star-forming galaxies at $z \gtrsim 1.8-2$,
    as well as to over-predict the number of low-luminosity galaxies at $z \lesssim 0.8$.
     }
   {Our study provides comprehensive galaxy number counts N(z) from galaxies with
    spectroscopic redshifts over a large redshift domain $0\leq z \le 5$,
    a solid basis for the measurement of volume-complete quantities.
    Magnitude-selected surveys identify a higher number of galaxies at $z>2$ than in colour-colour selected samples, 
    and we use the magnitude-selected VVDS to emphasize the large uncertainties associated to other surveys
    using colour or colour-colour selected samples. 
    Our results further demonstrate that semi-analytical models on dark matter 
    simulations have yet to find the right balance of physical processes and time-scales
    to properly reproduce a fundamental galaxy population property like the observed N(z).}


\authorrunning{Le F\`evre, O.}%, Cucciati, O., Cassata, P., et al.}
\titlerunning{VVDS: N(z) of magnitude limited samples to $z\sim4.5$}

   \maketitle

%
%________________________________________________________________

\section{Introduction}

Improving the understanding of galaxy evolution requires to study 
representative samples of galaxies at increasingly high redshifts.
The fast progress in the discovery of 
galaxies beyond $z \simeq 1-1.5$ has been primarily driven
by extensive deep spectroscopic redshift surveys.
Large samples with well controlled spectroscopic redshift measurements ensure
a robust foundation upon which the basic counting of galaxies at all epochs
can be performed, anchoring our understanding of galaxy evolution on firm grounds. 
High redshift spectroscopic surveys include samples purely magnitude-selected 
(CFRS: Le F\`evre et al. 1995, Lilly et al. 1995, K20: Cimatti et al. 2002, GDDS: Abraham et al. 2004, 
VVDS: Le F\`evre et al. 2005a, Garilli et al. 2008, Le F\`evre et al. 2013,
zCOSMOS: Lilly et al. 2007), 
colour-selected (DEEP2: Faber et al., 2007, GMASS: Cimatti et al., 2008), 
colour-colour selected including Lyman-break galaxies (LBGs), 
(e.g. Steidel et al. 1996,  Bouwens et al. 2007) 
and BzK  (Daddi et al. 2004), 
as well as Lyman-$\alpha$ emitters (LAE)  at increasingly high redshifts  
(e.g. Hu et al. 1998, Ouchi et al. 2008, Cassata et al. 2011), a non-exhaustive list.

%(Le F\`evre et al. 1995, Cimatti et al. 2002, Le F\`evre et al. 2005a, 
%Lilly et al. 2007, Garilli et al., 2008, Le F\`evre et al. 2013), by the effectiveness of colour selection 

Spectroscopic redshift surveys  form the basis from which 
the physical properties of the galaxy population and their evolution
are derived, most notably the luminosity function (LF), the
mass function (MF) and global quantities like the  star formation rate density (SFRD),
the stellar mass density, the merger rate, the clustering properties, as well as the
metallicity and dust content. These observed properties make the needed benchmark against which
to compare galaxy formation and evolution models combining major physical processes
like the merger rate, gas accretion, AGN feedback, or environment effects, just to
name a few. 

Because of the variety of selection functions, different surveys
do not sample the high redshift population in the same way,
and comparing different surveys while relating their galaxy populations to the 
global picture of galaxy formation and evolution is not straightforward. 
While the goal is to produce measurements relevant for the global population across all types of galaxies,
a complete census at all redshifts of all populations of galaxies from the highly star-forming
to the oldest most passive galaxies is yet to be produced, and we are faced with
several factors contributing to limiting the accuracy of our current census of the high redshift population. 

First and foremost, the representativeness of a sample is difficult to assess because 
we do not a priori know the population we are sampling. Observing strategies aim to sample
as completely as possible the entire galaxy population, but a range 
of different strategies have produced what is commonly referred to 
as a zoo of galaxy populations. High redshift surveys have produced
magnitude limited samples, 
%(CFRS: Le F\`evre et al. 1995, K20: Cimatti et al. 2002, GDDS: Abraham et al. 2004, 
%VVDS: Le F\`evre et al. 2005a, Garilli et al. 2008, zCOSMOS: Lilly et al. 2007), combined to
%colour selection (DEEP2: Faber et al., 2007, GMASS: Cimatti et al., 2008, VIPERS: Guzzo et al. 2013),
colour-colour selected samples (BzK, LBG),
%such as Lyman-break galaxies (LBGs) (e.g. Steidel et al. 2003), BzK galaxies
%(e.g. Daddi et al. 2004), 
Lyman-$\alpha$ emitter samples (LAE, Ouchi et al. 2008),
distant red galaxies (DRGs, van Dokkum et al., 2006), faint red galaxies (Franx et al.,
2003), as well as sub-mm galaxies (SMGs, Blain et al. 2002), IR selected LIRG/ULIRG (e.g. Oliver et al. 2010) and others. 
Relating and assembling these overlapping populations to build 
a complete census of the galaxy population at $z>1$ remains a
serious challenge. 

Another major difficulty is the estimate of the incompleteness
of each selection technique. Incompleteness can either be
built-in, with a specific selection purposely selecting only
a part of the global population, or be the result of the 
observations and data processing techniques, or a combination of both.
To estimate this incompleteness, surveys aim to securely identify
the redshift of all galaxies in their samples, with a good knowledge
of the selection function, and may rely on simulations
to compute the fraction of objects that are  a priori excluded
by observational strategies. 

Increasingly deeper multi-band photometry is used to compute
accurate photometric redshifts (see e.g. Wolf et al. 2003, Ilbert et al. 2006, Coupon et al. 2009, Ilbert et al. 2010,
Hidebrandt et al. 2012, Ilbert et al. 2013), and has become essential to identify galaxies up to the highest redshifts,
(e.g. Bouwens et al. 2010, McLure et al. 2010, Ellis et al. 2013). Photometry, by construction, 
goes deeper than spectroscopy on continuum-selected populations like the popular 
Lyman-break or Lyman-$\alpha$ drop-out technique.
However, spectroscopic follow-up often illustrates the difficulty to measure accurate redshifts of 
distant galaxies based on photometry alone, with degeneracies in redshift
estimates (e.g. Capak et al. 2011, Boone et al. 2011, Pirzkal et al. 2013), %the $z\sim5$ Mobasher, the $z\sim7$ presented at COSMOS conf...), 
and the large uncertainties associated to deriving global population
properties from photometric samples. Moreover, the accuracy of photometric redshifts
critically depends on the photometric bands used, their number and wavelength coverage, and, importantly, 
on the spectroscopic sample used to train them. 
Spectroscopic redshift measurements remain a necessary reference, especially at 
the higher redshift end which can be significantly
affected by catastrophic errors in photometric redshift measurements produced by photometric errors. 
%With about 15 times more galaxies at $z\sim1$ than at $z\sim3$ at a magnitude $i\sim24$, as reported
%later in this paper, even
%the best photometric redshifts with catastrophic error rates of about $\sim3$\% in photometric redshift
%measurement at $z\sim1$, may produce an uncertainty of up to $\sim$50\% in 
%estimating the number of galaxies at $z\sim3$.
Another limitation to current data-sets is the relatively
small field sizes which have been observed at $z>1$, making measurements 
sensitive to  large cosmic variance (e.g. Somerville, 2004). 
%Large and complete spectroscopic samples are therefore needed to overcome these limitations.
It is therefore necessary to proceed,
albeit with more difficulties, to assemble comprehensive samples of galaxies
with reliable spectroscopic redshifts.

%Making statements on evolution requires a detailed understanding on
%what population is actually observed, and the fraction of the population
%which is not observed. 

%Spectroscopic redshifts are required to eliminate most of the complex and hard to control
%completeness correction factors that need to be applied on colour or 
%narrow band selected galaxies samples at high redshifts. 

The knowledge of the redshift distribution N(z) is the basic information upon which all
of our current understanding of galaxy evolution is based upon, as to understand
galaxy evolution we must be able to identify  and count galaxies accurately, and it
is only with a proper accounting that we can envisage to perform detailed analysis
of the galaxy population and its evolution.  While the N(z) is a simple statistical description of
the galaxy population, it encodes all the physical processes acting 
to shape galaxies along cosmic time.
Some of the most important diagnostics of galaxy formation and evolution are
derived from galaxy counts: the SFRD evolution which can be derived from the LF and luminosity
density (LD), and the stellar mass density evolution derived from the mass function.
While these global properties contain a lot of information, their derivation
involves a number of uncertainties e.g. in computing rest-frame luminosities, stellar masses,
or effective volumes (Ilbert et al. 2004). 
As  redshift is increasing, uncertainties in computing LFs and MFs, and derivatives like LDs and SFRDs, 
are increasing significantly. 
%Placing secure constraints on the numbers of galaxies 
%requires a good handle on the density $\phi_*$,
%the characteristic luminosity $L_*$ and the slope $\alpha$
%of the LF.  
While a lot of progress has been
made in constraining the SFRD, measurements at
redshifts higher than $z\simeq1-1.5$ still show a significant spread.
The SFRD derived from the UV rest frame 
is relatively well constrained from $z\sim1-1.5$ to the present (e.g. 
Schiminovich et al., 2005; Tresse et al., 2007; Cucciati et al., 2012a). However, at $z\sim1.5-2$ 
measurements of the star formation rate show a factor 2-3 spread (Cucciati et al. 2012a, Behroozi et al. 2013),
in large part because finding galaxies in the 'redshift desert' is
difficult. Beyond $z\simeq2$ different measurements of 
the UV-derived SFRD are spread within factors $\sim3-10$ (e.g. Bouwens et al. 2007, Tresse et al. 2007,
Reddy et al. 2008, Cucciati et al. 2012a, Behroozi et al. 2013), illustrating the
incomplete knowledge of the galaxy population at these epochs, as well as the proper accounting
of all factors contributing to uncertainties (Poisson noise, completeness correction, fitting errors, cosmic variance,  etc.).  
Several factors contribute to limiting the accuracy
of LF/LD/SFRD measurements, most importantly the knowledge of the faint end
slope of the LF, the weak constraints on the bright
end of the LF because of the small volumes sampled, or dust correction
uncertainties, that are likely to combine to produce 
the discrepancy between the SFH derived from the LD and the SFH computed from the evolution
of the stellar mass density (e.g. Wilkins et al. 2008, Behroozi et al. 2013). 

The N(z), on the other hand, is a more straightforward observable to establish, involving less
uncertain steps, and, even if more complex to analyse, it must be reproduced
by any realistic model.
Projected number densities are straightforward to compare from one sample 
to another as they are based only on counts corrected for observational incompleteness. 
%Estimating LFs or MFs requires further steps with the computation of e.g. absolute magnitudes or
%stellar masses, as well as effective volumes
%which may vary somewhat depending on the method used,  
%or on the choice of statistical estimators, which may produce slightly different results, e.g.
%if the magnitude bias is not taken into account (Ilbert et al. 2004).
%The LFs and MFs are now well determined up to $z\sim1$ both from spectroscopic
%and photomeric redshift surveys (e.g. Cucciati et al. 2012), but  there is
%a large spread of measurements beyond $z\sim1$ in the SFR measurements (e.g. Behroozi et al. 2013, Fig. 2). 
%%photometric redshifts accuracies and catastrophic errors, or the depth 
%%and completness of spectroscopic surveys, lead to significant uncertainties resulting
%%in .
One of the main source of uncertainty in producing number counts is the accuracy and
reliability of the redshift measurements. 
Counts produced from photometric imaging surveys enable large samples of
galaxies with photometric redshifts to be assembled, reaching fainter magnitudes
than spectroscopic samples, although the photometric redshift accuracy is $\sim10^4$km/s at best. 
With an accuracy of $0.05 \times (1+z)$ down to $i_{AB}\sim25$
as is state of the art for multi-band photometric redshifts (see e.g. Ilbert et al. 2013), 
an error of $dz=0.2-0.3$ at $z\sim3$
translates to an error in absolute magnitude of 0.1-0.15, which may significantly
affect the accuracy of LFs computation. This is further complicated by catastrophic failures
(e.g. those galaxies with $|z_{spec}-z_{phot}|>>0.05 \times (1+z)$). 
%error in M_abs = 5log(1+z) - 5log(1+z+dz).
Spectroscopic redshift surveys are necessary to provide accurate redshifts,
even if they are targeting a subset of 
the galaxies detected in photometry and require to be corrected for incompleteness.

The knowledge of the N(z) is also an important element in studies of the
cosmological parameters of the world model,  like those using
weak lensing as a cosmology probe for which an accurate knowledge of the redshift distribution of the
background lensed population
is necessary to maintain a high accuracy and minimize biases in deriving
cosmological parameters with this technique (e.g. Fu et al., 2008)

In this paper we use the final data release of the VIMOS VLT Deep Survey (VVDS) 
as presented in 
Le F\`evre et al. (2013) to compute the redshift distribution N(z) of magnitude
limited samples. 
The VVDS final sample in the 0.61 deg$^2$ 0224-04 field is a purely 
$i-$band limited spectroscopic redshift survey, with a total of 10\,765 galaxies with
spectroscopic redshifts, going down to $i_{AB}=24.75$, and covering a redshift range $0<z<5$.

We summarize the properties of the VVDS sample in Section \ref{sample}. The 
redshift distributions and a parametric description
of the N(z) of $i-$band limited surveys at different
depths, as well as for $J-$band, $H-$band and $Ks-$band limited surveys,
are presented in Section \ref{sec_zdist}, and they are compared to
results from semi-analytic simulations. We present a complete census of
UV-selected galaxies in increasing redshift ranges in Section \ref{census} and compare them to 
galaxies selected using colour-colour criteria. We discuss these results in Section \ref{discuss},
and provide a summary in Section \ref{summary}.

%__________________________________________________________________

\section{VVDS sample and selection function}
\label{sample}

\subsection{Final sample of galaxies with spectroscopic redshifts from the completed VVDS}

The final VVDS data release is presented in Le F\`evre et al. (2013), augmenting the
previous data releases presented in Le F\`evre et al. (2005a) and Garilli et al. (2008), for a total
of $34\,594$ galaxies with spectroscopic redshifts obtained with VIMOS on the ESO-VLT (Le F\`evre et al. 2003)
\footnote{All VIMOS VLT Deep Survey data are publicly available on \texttt{http://cesam.lam.fr/vvds}}.
We use here $10\,044$ galaxies in the deep sample 
with $17.5 \leq I_{AB} \leq 24$, and $721$ in the ultra-deep sample with $23 \leq I_{AB} \leq 24.75$,
in the VVDS-02h field 
(at a location covered by XMM-LSS, SWIRE, CFHTLS-D1, GALEX, VLA, Herschel, among other ancillary observations), covering an area of
0.61 deg$^2$. An important component of the VVDS is the excellent associated photometry, with 
$BVRI$ from the CFH12K survey (Le F\`evre et al. 2004), subsequent deeper imaging in
$ugriz$ from the CFHT Legacy Survey (CFHTLS: Cuillandre et al. 2012), and near-infrared imaging in $YJHKs$ 
from the WIRDS survey (Bielby et al. 2012).

\subsection{VVDS selection function}
\label{sel_func}

The VVDS selection function is resulting from the target selection, the spatial and 
spectral sampling, and the completeness in redshift measurement. It is a key element in all
science analysis, and has therefore been extensively studied.
We refer the reader to the discussion in Le F\`evre et al. (2013) for a complete description,
and we provide here an overview of the main components of this selection function which
are needed to transform the observed galaxy counts N(z) into projected number densities
discussed in following sections. Of particular importance is the knowledge of the
survey incompleteness, to be able to produce completeness-corrected counts. We have performed
an extensive characterization of the survey selection function, and of the redshift
reliability flags, as described below.

The VVDS selection proceeds from a simple apparent $i-$band magnitude selection,
using the photometric catalogues produced by the CFH12K and CFHTLS imaging surveys.
With a density of galaxies in the parent photometric catalogue ranging from $\sim70\,000$ to $\sim85\,000$ gal/deg$^2$ at the
magnitude limits of the Deep and Ultra-Deep surveys resp., a 100\% sampling of the
galaxy population is impractical given current instrumentation, even with
the high multiplex of VIMOS (Le F\`evre et al. 2003). The VVDS has therefore observed a random selection
of galaxies, corresponding to about 25\% and 6.5\% of the full photometric parent
population in the Deep and UltraDeep surveys, respectively, and we call
this fraction the {\it Target Sampling Rate (TSR)}. After the spectroscopic observations
with VIMOS of this sub-sample and the data processing and redshift measurements, only a fraction of 
the spectroscopic targets provide a reliable redshift, and we call this the {\it Spectroscopic
Success Rate (SSR)}. The SSR is a function of both magnitude and redshift, and this
can be estimated using the redshift reliability flag scheme combined to photometric redshift estimates
from the extensive multi-band photometry, as originally presented in Ilbert et al. (2005), and derived in
Cucciati et al. (2012a) with a detailed presentation for the final VVDS data release in Le F\`evre et al. (2013). 

The redshift flag scheme used by the VVDS has evolved from the CFRS (Le F\`evre et al. 1995).
It relies on the comparison of redshifts independently measured by at least two persons
in the team, and reconciled to produce a final redshift measurement. The flags indicate
the probability for a redshift measurement to be right, and are therefore a reliability indicator
rather than a quality. Using the probabilistic nature of the redshift flags, it is then
possible to use galaxies with different flags, making a statistical correction
using the reliability of each flag, to infer global volume-representative quantities.
The VVDS flags 1 to 4 have associated reliability 51\%, 87\%, 98\% and 100\%,
respectively (Le F\`evre et al. 2013), with flag 0 for objects for which no redshift measurement could be
found, flag 9 for single emission line objects with $\sim80$\% reliability, and flag  1.5
introduced for the Ultra-Deep survey and corresponding to objects with a good match 
between the $z_{spec}$ and $z_{phot}$ within the photometric redshifts errors
(see Le F\`evre et al. 2013 for more details). The reliability estimates for each flag 
are very robust, as these have been derived from a number of independent observations
of the same objects, a result from the well known cancellation of individual biases of
observers when several independent measurements of the same variable are
performed. A total of 1\,263 
galaxies have been observed  independently twice within the VVDS (Le F\`evre et al. 2005a), as well as from
the MASSIV (Contini et al. 2012) and VIPERS (Guzzo et al. 2013) surveys, and data independently processed.
This gives a robust statistical baseline to
estimate the redshift reliability of the VVDS redshift flags, as described in Le F\`evre et al. (2013).

As part of the latest ultra-deep observations,
$241$ VVDS-Deep galaxies which ended-up with redshift flag 0, or flags 1 and 2 and redshifts $z > 1.4$, 
have been re-observed with
an exposure time of 16h, $\sim3.5\times$ longer than for the original VVDS-Deep observations, 
and with a larger wavelength domain $3600<\lambda<9300$\AA ~compared
to $5500<\lambda<9300$\AA ~for the VVDS--Deep. This enables to reduce the redshift degeneracies 
produced when only a few spectral features are present in a smaller wavelength domain.
The much deeper exposure time and broader wavelength coverage have led to a high success
rate in measuring the redshift of these re-observed galaxies, and hence help assess the redshift
distribution of the galaxies with lower redshift reliability flags 1 or 2, 
or failed redshift measurements (flags 0) of 
the VVDS--Deep sample. Indeed the success rate (flag 2, 3, 4, 9) of re-observed flag 0 is 85\%,
for flag 1 it is 86\%, and for flag 2 88\%. Using this re-observed sample, we have 
computed the photometric sampling rate, the target sampling rate, and the spectroscopic success rate (SSR),
by redshift range and by flag category, leading to a robust understanding of the incompleteness
of the sample, and of the redshift distribution of the failed population.
In addition, we have used the extensive $u*,g',r',i',z',B,V,R,I,J,H,$ and $Ks$
photometry in the VVDS-02 field (Le F\`evre et al. 2004, McCracken et al. 2003, CFHTLS-D1, Bielby et al., 2012) 
iterating from Ilbert et al. (2006) to compute accurate photometric redshifts 
which have then been compared to the observed spectroscopic redshifts. 
The results of this analysis are presented in Le F\`evre et al. (2013), which shows
an excellent agreement between $z_{spec}$ and $z_{phot}$, with $dz=0.04 \times (1+z)$
and a catastrophic failure rate of a few percent.  

The complete understanding of the sample selection function summarized in this section, 
and detailed in Le F\`evre et al. (2013), is used in the following sections to
study complete magnitude limited samples. The galaxy samples used in this paper are free from
star or broad-line AGN contamination, as the VVDS spectroscopy easily enables to identify these objects, 
in a more straightforward way than from photometric samples.

\section{Redshift distribution of $i-$, $J-$, $H-$, and $Ks-$band magnitude limited samples}
\label{sec_zdist}

\subsection{Method}

The average $N(z)$ provides a simple but important 
description of the galaxy population and its evolution, combining the luminosity distribution 
of all types of galaxies at a fixed redshift. 
%The proper knowledge of
%this information is also important to a number of other studies, e.g. for weak lensing studies
%which need to assume a redshift distribution of background sources. 
In this section we
derive the observed N(z) for $i-$band as well as $J$, $H$, and $Ks-$band selected samples to the unprecedented
depth $i_{AB}=24.75$, $J_{AB}=23$, $H_{AB}=22.5$, $Ks_{AB}=22$, and present a parametrisation to 
describe these distributions analytically.  

We have corrected our sample for the selection function counting each galaxy with the 
following weight $W_{gal,i}$:
\begin{equation}
W_{gal,i}= 1 / _{TSR} \times 1 / _{SSR} \times 1 / _{PSR} \times 1 / w_{129}
\end{equation}
The Target Sampling Rate $TSR$ and Spectroscopic Success Rate $SSR$ are as defined
in Section \ref{sel_func}.
The $PSR$ is the Photometric Sampling Rate, i.e. the ratio of objects in the photometric
catalogue used to select spectroscopic targets, over those in the parent photometric sample, in a given magnitude bin;
this applies only to the Ultra-Deep survey as some galaxies with $23 \leq i_{AB} \leq 24.75$
have already been observed in the Deep survey, and PSR=1 for the Deep survey.
The weight $w_{129}$ is the ratio of the number of galaxies with the lower 
reliability flags 1, 2 and single emission line flag 9 over the
number of galaxies with redshifts in the same redshift bin, using a combination of photometric redshifts and spectroscopic redshifts
from the Ultra-Deep sample.
The behaviour of these weights is extensively described in Le F\`evre et al. (2013),
and have been further described and used in Cucciati et al. (2012a).
We note that $w_{129}$  may be overestimated at $z>2.5$ because of a 
higher catastrophic failure rate of the photometric redshifts used to compute
this weight, resulting in an artificial lowering of the counts once this correction has been applied;
the effect of this uncertainty is further discussed in Section \ref{i24}.

Applying the global weight $W_{gal,i}$ on each galaxy 
then provides corrected number counts, as described below.

\subsection{Cosmic variance}
\label{cosv}

We use the prescription detailed in Moster et al. (2011) 
to derive cosmic variance estimates for each of our samples.
The main contribution of cosmic variance on galaxy numbers uncertainties is the
volume probed, which depends on the area of the survey and the
redshift range spanned. Because of the galaxy to dark matter bias,
Moster et al. (2011) emphasize that cosmic variance also depends 
on the stellar mass range of galaxies observed, and we have used stellar mass estimates
based on the code Le Phare as described in Ilbert et al. (2010, 2013).
For each sample we then followed Moster's et al. (2011) cookbook, using 
a field size of 0.61 deg$^2$, a mean redshift $\bar{z}$ and redshift bin size $\Delta z$
as well as a mass range adapted to each sample.
Given the magnitude selection of the VVDS samples, the mass range is increasingly
probing higher masses at higher redshifts, and hence the cosmic variance is
globally increasing with redshift and ranges from about 6\% in the best cases
to $\sim20-30$\% at the highest redshifts, in $\Delta z=0.5$. 
The cosmic variance estimates are provided as a function of redshift
for the redshift distribution N(z) presented in this section \ref{sec_zdist}. 
For the projected number counts described in Section \ref{census}, 
the cosmic variance has been computed in the corresponding redshift bins,
and added in quadrature to the number counts Poisson errors, to produce 
total count uncertainties.

%\subsection{$17.0 \leq i_{AB} \leq 22.5$}

\subsection{$17.5 \leq i_{AB} \leq 24$}
\label{i24}

The redshift distribution in this magnitude range
has been presented in \cite{olf2} from the first epoch
VVDS--Deep sample. Here we are adding the measurements 
from the second epoch 'Deep' observations described in Le F\`evre et al., (2013). 
The total sample of galaxies
with spectroscopic redshifts and with a flag larger or equal to 1 and $17.5 \leq i_{AB} \leq 24$ is 10\,044. 

%The very deep re-observations of the failed redshift measurements
%in the VVDS--Deep sample described in Section \ref{sel_func}
%enable to correct the VVDS--Deep sample for completeness. 
%We were able to obtain reliable redshift
%measurements for 85\% of this re-observed sample. 
The original VVDS--Deep was $\simeq80$\% complete in redshift measurements (those galaxies 
with redshift reliability flags 2, 3, 4 and 9), and
the re-observed sample taken from the 20\% incompleteness now has reliable
redshifts for $\simeq85$\% of this originally failed (flag 0)
or low reliability (flag 1) sample. Although only a fraction, %241 galaxies over 528 with flags 1 and 2),
and not all, of this low reliability sample have been re-observed, this nevertheless allows to
statistically correct the sample to a spectroscopic
completeness of the $i_{AB} \leq 24$ sample equivalent to having 97\% of the sample
with a reliable redshift. 
This produces a spectroscopic redshift distribution free of bias, as
the remaining 3\% of the sample is not expected to modify the shape of the N(z).

\begin{figure}
   \centering
  \includegraphics[width=8cm]{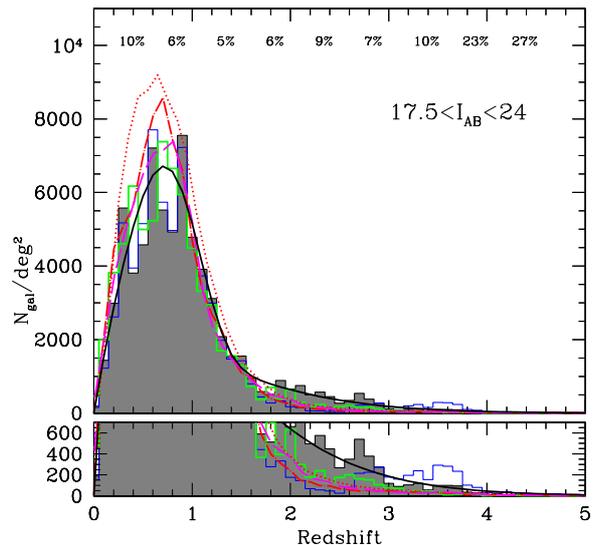}
      \caption{Spectroscopic redshift distribution N(z) (number of galaxies per square degree) of 
               galaxies with $17.5 \leq i_{AB} \leq 24$, from the
               VVDS--Deep sample before
%using all flags larger or equal to 1.5 
	       (blue histogram), and after correction for 
               the selection function 
               (filled histogram). The continuous black line is the best fit using 
               equation \ref{eq1}. The N(z) from the De Lucia and Blaizot (2007) SAM based on the Millennium simulation
               using the WMAP1 cosmology is shown as the dotted red line,
               the SAM based on the Millennium-WMAP3 as the dot-dash red line (Wang et al., 2008), and
               the latest Millenium-WMAP1 SAM as the dashed magenta line (Henriques et al., 2012). 
		The open green histogram is the N(z) derived from the updated v2.0 photometric redshift sample from 
               Ilbert et al. (2009), including UltraVista data, on 1.73 deg$^2$ in the COSMOS field. 
               Estimates of cosmic variance are listed on the top of the plot. 
              }
         \label{nz_deep}
   \end{figure}

The corrected and uncorrected redshift distributions for this sample using
the selection function obtained in Section \ref{sel_func} are shown in
Figure \ref{nz_deep}. We see very well the effect of the selection function correction,
which fills-in the 'redshift desert' at $1.5<z<2.7$ as produced by the LRRED grism
observations, where most of the failed redshift
measurements in the original VVDS--Deep  originated from, as had
been anticipated (Le F\`evre et al. 2005a, Ilbert et al. 2005, Tresse et al. 2007). 

As a consequence, the corrected N(z) has a mean redshift of $\bar{z}=0.92$ with a broad 
distribution with $\sigma=0.67$. Because of the selection function correction, 
the mean redshift is significantly different from the
value $\bar{z}=0.78$ without correction that was quoted in Le F\`evre et al. (2005a).

One can note that the corrected N(z) at $z>2.5$ is significantly lower than
the observed N(z). This is the result of applying the weight $w_{129}$,
which has a value larger than 1 at these redshifts. While this translates that
a number of low reliability flag 1 must be at lower redshifts than indicated
by their measurements, it also includes
increasing uncertainties and catastrophic failures of photometric redshifts
which lead to increase  $w_{129}$, and hence decrease the N(z) once
this weight is applied. The corrected N(z) at $z>2.5$ for this magnitude
range is therefore likely to be an under-estimate. This notwithstanding,
the fraction of galaxies with $z \geq 2$ at this depth is 8.2\%.

The VVDS N(z) from spectroscopic redshifts is compared in Figure \ref{nz_deep} to the N(z) 
derived from photometric redshifts in the COSMOS field, i.e. in a field completely independent from the VVDS-02h Deep field.
We use a new version v2.0 of the $i-$band selected photo-z catalogue
from Ilbert et al. (2009) which incorporates the new UltraVISTA data
(McCracken et al. 2012, Ilbert et al. 2013), calibrated
on spectroscopic redshift samples independent from the VVDS redshifts.
The agreement is quite impressive up to $z\simeq2$, while the high redshift
tail at $z>2$ is less populated in the photometric N(z).

We have made an attempt to fit the observed N(z) with analytic functions
proposed in the literature (e.g. Wilson et al., 2001; van Waerbeke, 2001; Schrabback et al., 2010).
However, as these studies did not deal with redshifts $z>1.5$, we found that these 
functions are not capable to reproduce the high redshift 
tail observed at $z>1.5$. We therefore slightly modified the form proposed
by Schrabback et al. (2010) to 
%\[
\begin{equation}
N(z)=A \times ( z/z_0 )^{\alpha} \left[ exp \left( - ( z/z_0 ) ^{\beta} \right)  + B \times exp \left( - ( (z+z_t)/z_0 ) ^ {\gamma}  \right) \right]
\label{eq1}
\end{equation}
%\]
%c1 * ( ( x/c3 )  ** c2 ) * ( exp ( - ( x/c3 ) ** c4 )  + c5 * exp ( - ( (x+c7)/c3 ) ** c6 ) )
which better takes into account the high redshift tail by the addition of a second exponential term. 
The result of the best fit using this formula is shown by the continuous line in Figure \ref{nz_deep}, and the best 
fit parameters are listed in Table \ref{param_nz}.
%This produces a new reference for the N(z) of very deep galaxy samples.

   \begin{table*}
      \caption[]{Best fit parameters for redshift distributions N(z) (following equation \ref{eq1})}
      \[
 %        \begin{array}{p{0.5\linewidth}ll}
        \begin{array}{lcrrrrrrr}
           \hline
            \noalign{\smallskip}
Sample  & \bar{z} & A & \alpha & z_0 & \beta & B & \gamma & z_t \\
            \noalign{\smallskip}
            \hline
            \noalign{\smallskip}
17.5\leq I_{AB} \leq 24    & 0.92 & 9\,173\pm228 & 0.99\pm0.03 & 1.03\pm0.01 & 3.88\pm0.14 & 24\pm10 & 1.20\pm0.13 & 2.9\pm1.2 \\
17.5\leq I_{AB} \leq 24.75 & 1.15 & 8\,888\pm216 & 1.06\pm0.02 & 1.07\pm0.01 & 6.19\pm0.27 & 36.5\pm27 & 1.17\pm0.10 & 2.7\pm1.1 \\
23.0\leq I_{AB} \leq 24.75 & 1.38 & 3\,519\pm108 & 2.11\pm0.10 & 2.10\pm0.02 & 2.20\pm0.06 & 202\pm50 & 2.85\pm0.30 & 2.3\pm0.7 \\
J_{AB} \leq 23.0 & 0.95  & 2\,100\pm260 & 0.95\pm0.08 & 1.13\pm0.02 & 4.80\pm0.16 & 320\pm47 & 1.37\pm0.21 & 2.9\pm1.5 \\
H_{AB} \leq 22.5 & 0.94  & 2\,200\pm275 & 0.94\pm0.08 & 1.15\pm0.02 & 5.00\pm0.29 & 205\pm53 & 1.35\pm0.17 & 2.9\pm1.4 \\
Ks_{AB} \leq 22.0 & 0.88  & 2\,600\pm110 & 0.98\pm0.08 & 1.18\pm0.01 & 6.90\pm0.35 & 120\pm10 & 1.38\pm0.25 & 2.9\pm1.7 \\
            \noalign{\smallskip}
            \hline
         \end{array}
      \]
%\begin{list}{}{}
%\item[$^{\mathrm{a}}$] This is footnote a
%\end{list}
\label{param_nz}
   \end{table*}

\subsection{$23.0 \leq i_{AB} \leq 24.75$}

\begin{figure}
   \centering
   \includegraphics[width=8cm]{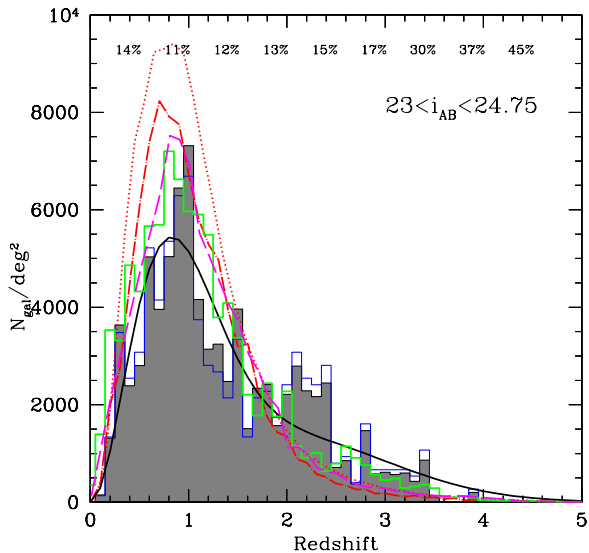}
      \caption{Spectroscopic redshift distribution N(z) (number of galaxies per square degree) of 
               galaxies with $23 \leq i_{AB} \leq 24.75$,
               before %correction using all flags larger or equal to 2 
               (blue histogram), and after correction using the selection function 
               obtained in Section \ref{sel_func} (filled histogram).  The best fit using equation \ref{eq1} 
               is shown as the continuous black line, and best fit 
               values are given in Table \ref{param_nz}.  
		The N(z) from the De Lucia and Blaizot (2007) SAM based on the Millennium simulation
               using the WMAP1 cosmology is shown as the dotted red line,
               the SAM based on the Millennium-WMAP3 as the dot-dash red line (Wang et al., 2008), and
               the latest Millenium-WMAP1 SAM as the dashed magenta line (Henriques et al., 2012). 
		The open green histogram is the N(z) derived from the updated v2.0 photometric redshift sample from 
               Ilbert et al. (2009), including UltraVista data, on 1.73 deg$^2$ in the COSMOS field. 
               Estimates of cosmic variance are listed on the top of the plot. 
              }
         \label{nz_udeep}
   \end{figure}

The Ultra-Deep sample observations cover a large observed wavelength
range $3600<\lambda<9300$\AA. There are always 
spectral features available for redshift measurement from $z=0$ to $z\sim6$, and 
therefore the Ultra--Deep sample does not present a 'redshift desert'. This enables
to derive a redshift distribution N(z) with a high success rate at these magnitudes, 
the only $i-$band selected spectroscopic sample at this depth so far. 
Using the total sample of 622 galaxies with $23.0 \leq i_{AB} \leq 24.75$
and with flags  $\geq1.5$ (see Le F\`evre et al. 2013), we find that 
the mean redshift is $\bar{z}=1.38$ with a broad distribution with $\sigma=0.81$,
indicating that the sample is probing the luminosity function sufficiently 
fainter than the mean luminosity $L_*$ when going to redshifts higher than $z\sim1.5$,
producing the observed increase of galaxies numbers in the high redshift tail. In this magnitude range
23.5\% of galaxies are at $z \geq 2$.
The result of the best fit using Equation-\ref{eq1} is shown in Figure \ref{nz_udeep}, 
and the best fit parameters are listed in Table \ref{param_nz}.

We compare the VVDS N(z) from spectroscopic redshifts  to the N(z) 
derived from photometric redshifts in the COSMOS field (Ilbert et al. 2009), as shown
in Figure \ref{nz_udeep}.
At these faint magnitudes the agreement remains excellent although with 20-25\% more
galaxies below $z\simeq1.5$, and $\sim25$\%
less galaxies in the high redshift
tail at $z>2$, in the photometric redshift distribution compared to the VVDS. 

\begin{figure}
   \centering
   \includegraphics[width=8cm]{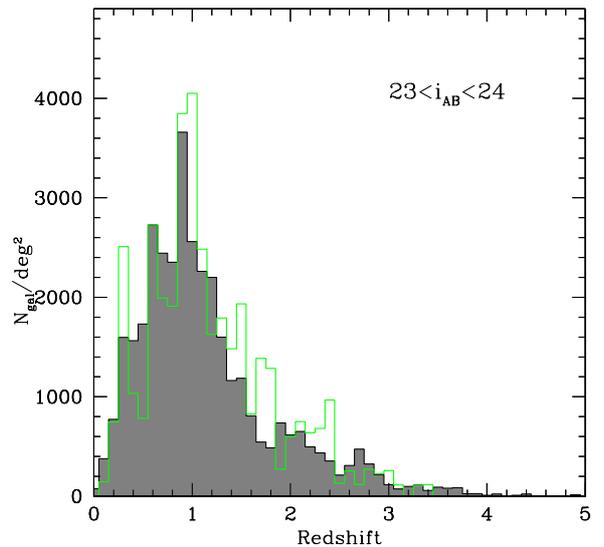}
      \caption{Comparison of the spectroscopic redshift distribution N(z) (number of galaxies per square degree) of 
               galaxies with $23 \leq i_{AB} \leq 24$, as derived from the VVDS Deep sample (filled histogram)
               and from the VVDS Ultra-Deep sample (open green histogram).
               The two distribution are drawn from independent samples, and agree very well, within
               statistical errors.
              }
         \label{nz_23_24}
   \end{figure}

The Deep and Ultra-Deep samples overlap in magnitude range over $23 \leq i_{AB} \leq 24$,
with different galaxy samples observed with different instrument set-ups,
and therefore provide two independent samples that can be compared to evaluate the
robustness of the procedures applied to measure the redshift distributions.
The N(z) for these two samples are shown in Figure \ref{nz_23_24}. These two distributions
agree very well, considering that
the N(z) for the Ultra-Deep comes from an area 4.3 times smaller than for the 
VVDS and hence has a larger intrinsic scatter. 
%cosmic variance here does not apply as this is the same field and the 
% same volume is sampled

\subsection{$17.5 \leq i_{AB} \leq 24.75$}

\begin{figure*}
   \centering
   \includegraphics[width=15cm]{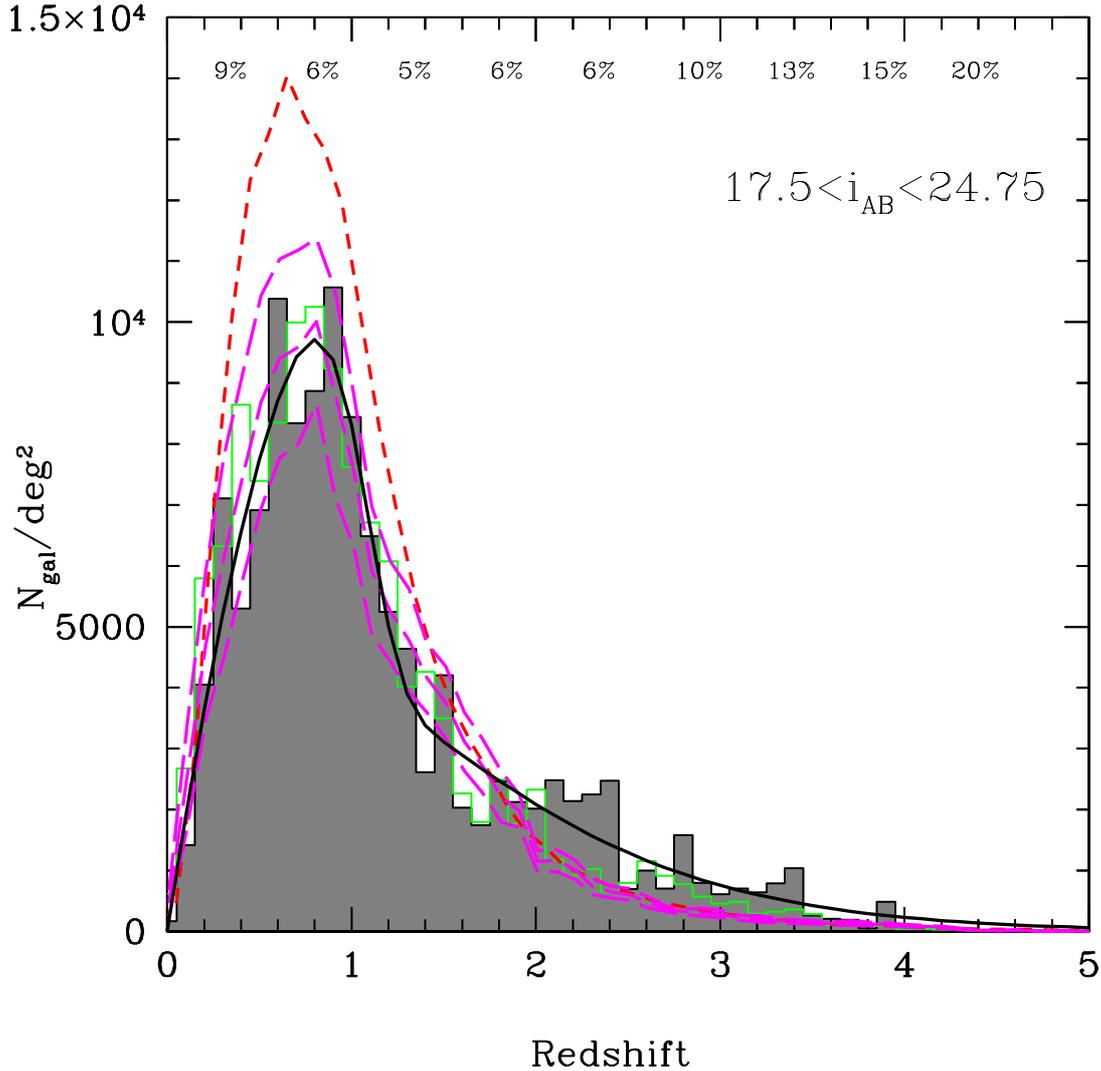}
      \caption{Spectroscopic redshift distribution N(z) (number of galaxies per square degree) of 
               galaxies with $17.5 \leq i_{AB} \leq 24.75$,
               using the VVDS--Deep ($10\,044$ galaxies) and VVDS-UltraDeep samples (721 galaxies),
               in 0.61 deg$^2$,
               corrected for the selection function obtained in Section \ref{sel_func}. 
               The continuous black line is the best fit using equation \ref{eq1},
               with best fit values given in Table \ref{param_nz}. 
               The N(z) from the De Lucia and Blaizot (2007) SAM based on the Millennium simulation
               using the WMAP1 cosmology is shown as the dotted red line
               and the SAM based on the latest Millennium-WMAP1 as the dashed magenta lines (Henriques et al., 2012)
               representing the mean and $\pm1\sigma$ values from 24 mocks.
		The open green histogram is the N(z) derived from the updated v2.0 photometric redshift sample from 
               Ilbert et al. (2009), including UltraVista data, on 1.73 deg$^2$ in the COSMOS field. 
               Estimates of cosmic variance are listed on the top of the plot. 
              }
         \label{nz_tot}
   \end{figure*}

Combining the VVDS--Deep and the VVDS--UltraDeep samples we are able to
compute for the first time the $N(z_{spec})$ from a  magnitude-limited spectroscopic sample
with $17.5 \leq i_{AB} \leq 24.75$. We have used the weighting scheme described above
applying the TSR, SSR, PSR, and $w_{129}$ weights computed for the VVDS-UltraDeep survey
to each galaxy. The result is shown in Figure \ref{nz_tot}. The mean redshift is 
$\bar{z}=1.15$, with a significant high redshift tail going up to $z\simeq5$. 
The fraction of galaxies with $z \geq 2$ is 17.1\%
The result of the best fit is also shown in Figure \ref{nz_tot}, with the best fit 
parameters listed in Table \ref{param_nz}.

The photometric redshift distribution derived from photometric redshifts in the COSMOS
field (Ilbert et al. 2009) is shown in Figure \ref{nz_tot}. As noticed in
previous sections, the agreement is excellent up to $z\sim2$, while above this redshift
the N($z_{phot}$) counts are lower by 25-30\%.

\subsection{J, H, and Ks selected samples}

%$20 \leq J_{AB} \leq 23$, $19.8 \leq H_{AB} \leq 22.5$, $19.5 \leq K_{AB} \leq 22.0$

\begin{figure}
   \centering
   \includegraphics[width=8cm]{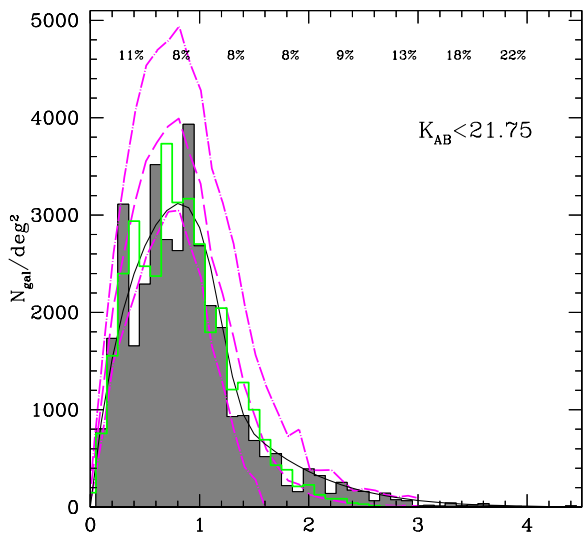}
      \caption{Spectroscopic redshift distribution N(z) (number of galaxies per square degree) of 
               galaxies with $Ks_{AB} \leq 21.75$ to ensure $\sim100$\% completness,
               after correction for the selection function obtained in Section \ref{sel_func}. 
		The best fit using equation \ref{eq1} is shown as the 
               continuous black line, and the best fit 
               parameters are listed in Table \ref{param_nz}.
               The $Ks_{AB} \leq 21.75$ redshift distribution from the SAM model of Henriques et al. (2012) 
		on the millennium simulation
               is shown as the dashed magenta line, with $\pm1\sigma$ values indicated as the dot-dash lines.    
	       The open green histogram is the N(z) derived from the photometric redshift sample of 
               Ilbert et al. (2013) from 1.48 deg$^2$ in the COSMOS-UltraVista field. 
               Estimates of cosmic variance are listed on the top of the plot      
              }
         \label{nz_k}
   \end{figure}

From our  sample we are also able to derive magnitude limited samples in each of the 
$J, H$ and $Ks$ bands. In these bands the VVDS spectroscopic samples allow to identify
samples nearly 100\% complete in redshift success rate down to $J_{AB}=23$, $H_{AB}=22.5$ and $Ks_{AB}=22$
(Le F\`evre et al. 2013). 
There are 2023, 1679, 1457 galaxies, and
the mean redshifts are 0.94, 0.92, and 0.88 for the $J$, $H$, and $Ks-$ limited samples at these
depths, respectively.
We present the redshift distribution down to $Ks_{AB}=22$ in Figure \ref{nz_k}. 
Best fit values using equation \ref{eq1} are given in Table \ref{param_nz}.

%   \begin{figure}
%   \centering
%   \includegraphics[width=8cm]{/home/olefevre/vvds/science/survey/udeep/jul09/histz_JHK_fit.ps}
%      \caption{Spectroscopic redshift distribution N(z) (number of galaxies per square degree) of 
%               galaxies with $J_{AB} \leq 23$, $H_{AB} \leq 22.5$, and $K_{AB} \leq 22$,
%               after correction for the selection function obtained in Section \ref{sel_func}. 
%		The best fit using equation \ref{eq1} is shown as the 
%               continuous line in each panel, and the best fit 
%               parameters are listed in Table \ref{param_nz}.
%               The $K_{AB} \leq 22$ redshift distribution from the SAM model of Wang et al. (2008) on the millennium simulation
%               is shown as the dotted line            
%              }
%         \label{nz_k}
%   \end{figure}

\subsection{Comparison with the Millennium SAM simulations}

The observed N(z), as a function of the magnitude limit of a sample,
offers the possibility to test  semi-analytic models (SAM).
We have used SAM implemented on the Millennium Simulation (Springel et al. 2005, using the Millenium database: Lemson \& the VIRGO
consortium 2006) 
from de Lucia and Blaizot (2007) and the Munich SAM (Guo et al., 2011). 
While the Millennium simulation is based on WMAP1 cosmology (Spergel et al., 2003), it
can be scaled to other cosmologies using the technique of Angulo \& White (2010). We have used the 
WMAP1 and WMAP3 cosmology versions of the Munich SAM produced by Wang et al. (2008). 
In addition, we have used the latest implementation of the Munich SAM
with WMAP1 cosmology as described by Guo et al. (2013). As detailed in Guo 
et al. (2013), the WMAP1 and WMAP3 versions of the SAM are expected to represent
the extremes of the galaxy population as the SAM based on WMAP7 (Komatsu et al., 2011) seem to converge in between these
two extremes (Guo et al., 2013), so we expect that the WMAP7 version of the SAM will fall in between
the WMAP1 and WMAP3 that we describe below. The effects on the models of going to the slightly different cosmology
from the Planck results (Planck collaboration 2013) remains to be calculated.
We have compared the SAM results when using Bruzual and Charlot (2003, BC03) or the Maraston (2005, M05) 
galaxy stellar population synthesis model libraries. We find that the M05 N(z) in the simulations is always giving
lower counts at $z>2$ than BC03, hence further increasing the differences with the observed VVDS N(z),
while counts for $z<2$ are closely comparable.
As our results and discussion are further reinforced when using M05 we have opted rather to compared to BC03
Millenium models in the following, so that our conclusions remain independent of the choice of libraries.

The N(z) derived from the Millennium SAM for $i-$band selected samples is shown in Figures \ref{nz_deep}, \ref{nz_udeep}  
and \ref{nz_tot}, expanding from the work of de la Torre et al. (2011).
Two main discrepancies between current semi-analytic model predictions
and observations are evident: the simulations predict too many faint galaxies at low redshifts,
and, on the contrary, the simulations do not contain enough galaxies beyond $z>2$. 
It can be seen from Figures \ref{nz_deep}, \ref{nz_udeep}  
and \ref{nz_tot} that the discrepancy between the observed and model N(z) is 
increasing with magnitude. 
For the sample with $17.5 \leq i_{AB} \leq 24$ all
models used are about $\sim40$\% higher than the observed N(z), 
which is statistically significant at the $3\sigma$ level when taking into account 
the Poisson and cosmic variance on the data side, and the $1\sigma$ values on 
the simulated N(z) as derived from 24 realisations. 
Between redshifts $2.5<z<3.5$, the situation is reversed: the observed N(z)
is $\sim2.5\times$ larger than the highest simulated set, significant at the $4\sigma$ level.
Models for the fainter sample with $23 \leq i_{AB} \leq 24.75$
present excess in counts by a factor $\sim1.8$ at $0.7< z < 0.9$, and counts lower by a 
factor $\sim2-3$ at $z\sim2-3$, the discrepancy being higher (lower) for WMAP1 than for WMAP3 cosmology
at lower (higher) redshifts. When taking into account cosmic variance, these differences are
also statistically significant at the $3-4\sigma$ level.
A similar behaviour is observed for the $Ks$ band, with more galaxies predicted by the model
at the lower redshifts than observed. It is also evident that the model N(z) for a $Ks_{AB} \leq 21.75$ sample 
has lower counts than observed at $z>2$, as shown in Figure \ref{nz_k}.
This is further discussed in Section \ref{discuss}.

\section{A complete census of star-forming galaxies with $1.4 < z \leq 4.5$}
\label{census}

%\subsection{Galaxy Counts from magnitude limited samples in u, g, r, i, J, and K bands}

%Using the completeness-corrected spectroscopic redshift distribution N(z), 
%we can derive the projected number density of galaxies selected with
%$17.5 \leq i_{AB} \leq 24.75$ in several redshift bins, as presented in
%Figure \ref{counts_i}.

%From the 

%   \begin{figure}
%   \centering
%   \includegraphics[width=8cm]{empty.eps}
%      \caption{Projected number counts per square arcmin of galaxies
%       with $17.5 \leq i_{AB} \leq 24.75$ in several redshift bins}
%         \label{counts_i}
%   \end{figure}

\subsection{Projected galaxy number counts in selected redshift domains}

In this section we aim to obtain accurate counts of galaxies in
increasing redshift ranges up to $z=4.5$, as these are the basic input to the
computation of statistical estimators like Luminosity Functions and Star Formation Rate Density,
or mass functions and stellar mass density.

Here we take advantage of the VVDS   spectroscopic redshifts sample,
with low levels of incompleteness, to measure projected number
counts.  All the counts presented below are corrected %for the PSR, SSR, and TSR
as described in Section \ref{sel_func} and in Le F\`evre et al. (2013). 
All integrated count errors include Poisson noise
as well as cosmic variance noise as described in Section \ref{cosv}.
We are concentrating on high redshifts beyond $z=1.4$ in order to match
popular photometric selection techniques: the BzK-selection in $1.4 \leq z \leq 2.5$,
and the LBG-selection in $2.7 \leq z \leq 3.4$ and $3.4 \leq z \leq 4.5$.  
The redshift range $0 < z \leq 1.4$ is extensively discussed in Cucciati et al. (2012a).

All the VVDS counts are presented in Table \ref{counts_zint}. 

%\subsection{Galaxies with $0 < z \leq 1.4$}
%
%The counts in this redshift range are presented in Cucciati et al. (2012), they
%include the weighting corrections as presented in Le F\`evre et al. (2013).
%%Using the VVDS-Deep and VVDS-UltraDeep final data as presented in Le F\`evre et al. (2013), 
%%a better understanding 
%%of the incompleteness of the first epoch VVDS data in computing the UV-LF .
%%Combined with a larger sample hence improved statistics, and a deeper sample 
%%to constrain the slope $\alpha$ to fainter flux.
%The evolution of the UV LF over the 
%redshift range $0<z<1.4$, and derived the evolution of the SFRD, improving on
%our earlier measurements (Ilbert et al. 2005, Tresse et al. 2007).
%The projected counts are listed in Table \ref{}.
%We find that... in agreement with...

\subsection{Galaxies with $1.4 \leq z \leq 2.5$}
\label{z2}

This redshift range is important as the peak of star formation seems to
be occurring at this epoch. However, the secure identification of galaxies at
these redshifts is complicated because of the lack of spectral features
in the visible domain, making it difficult to measure spectroscopic
redshifts (the so-called 'redshift desert'). 

\begin{figure}
   \centering
   \includegraphics[width=8cm]{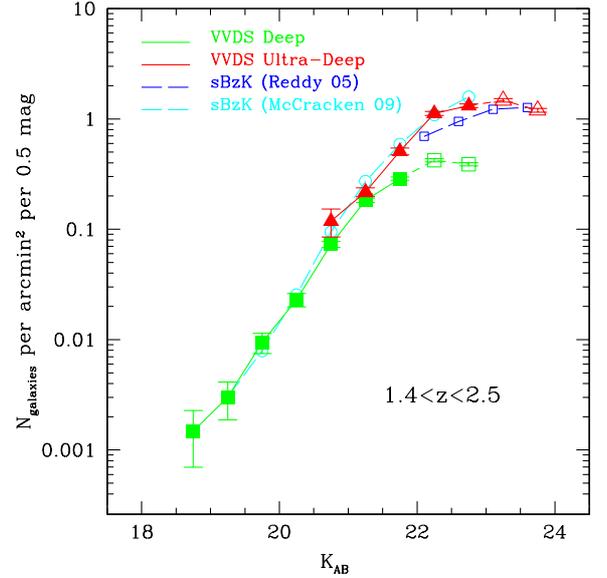}
      \caption{Projected galaxy numbers per square arc-minute as a function of
       $Ks_{AB}$ magnitude for galaxies with spectroscopic redshifts $1.4 \leq z_{spec} \leq 2.5$, from the 
       VVDS--Deep (filled squares), the VVDS--UltraDeep (triangles), compared to
       data for galaxies selected from the BzK technique (empty squares: Reddy et al. (2005); 
       empty circles: McCracken et al. (2009)). For the VVDS--Deep and Ultra-Deep, the 
       open symbols and dashed lines indicate the magnitudes for which the samples are incomplete. }
         \label{dens_bzk}
   \end{figure}

The spectroscopic redshifts of 501 galaxies with flags 2, 3, 4 and 9 have been successfully
measured in this range, with
another 354 with flag 1. 
The projected number counts are presented in Figure \ref{dens_bzk}. We 
find $2.07\pm0.12$ gal/arcmin$^2$ with $1.4 \leq z \leq 2.5$ down to magnitude $Ks_{AB}=22.5$,
and  $1.17\pm0.08$ gal/arcmin$^2$ with $22 \leq Ks_{AB} \leq 22.5$ (the errors are including cosmic variance). 
We find a good agreement between the counts from the VVDS--Deep and  Ultra-Deep samples
in the magnitude range of overlap. 

We find comparable number counts to the BzK selected sample
of McCracken et al. (2009), as well as those of Kong et al. (2006),
Blanc et al. (2008), and Daddi et al. (2004). However, our counts are about twice larger than those of 
Reddy et al. (2005) who find about 0.6 gal/arcmin$^2$ down to $Ks_{AB}\simeq22$, 
and our counts are still 25 to 40\% higher than theirs at the faint end $22 \leq Ks_{AB} \leq 23$ even 
though our $Ks-$band counts 
start to be incomplete fainter than $Ks_{AB}=22$ as a result of the VVDS $i-$band selection. 

\begin{figure*}
   \centering
   \includegraphics[width=14cm]{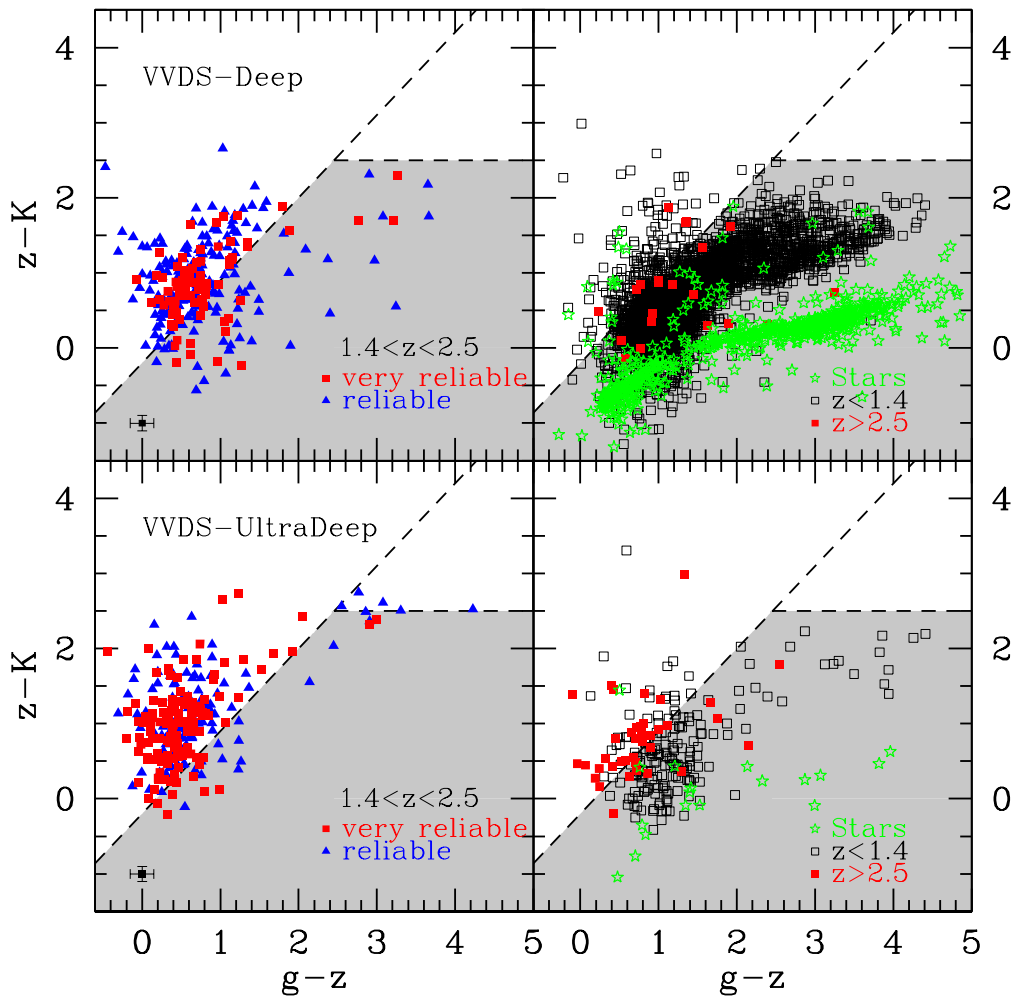}
      \caption{Efficiency of the gzK selection criterion to
               identify galaxies at $1.4 \leq z \leq 2.5$ compared to
               $i-$band magnitude selection from the VVDS Deep sample {\it (top)} and
               from the Ultra-Deep sample {\it (bottom)}: 
               {\it (left)} galaxies with spectroscopic redshifts
               in $1.4 \leq z \leq 2.5$ (flags 2: blue triangles, flags 3, 4, 9: red squares). 
               Average colour errors are indicated on the bottom left of the top and bottom  panels.
               Galaxies in the shaded area would be
               excluded from the gzK colour selection as being at redshifts outside $1.4 \leq z \leq 2.5$,
               although their spectroscopic redshifts indicate otherwise.
               Using the most reliable flag 3,4, and 9 magnitude-selected spectroscopic redshift sample,
               we find that $\sim10$\%
               of the total number of galaxies with a correct spectroscopic redshift would not be selected using
               gzK selection criterion (\ref{sgzk}).
               {\it (right)} galaxies with spectroscopic redshifts $z < 1.4$ (black open squares) or $z > 2.5$
               (red filled squares), and stars (green starred symbols).
               Galaxies outside the shaded area would be wrongly selected by the gzK selection as being at
               $1.4 \leq z \leq 2.5$, using only the most reliable flags 3, 4 and 9 in the VVDS-UltraDeep
               they represent 23\% (resp. 34\%) of the  total number of galaxies with the 
               correct spectroscopic redshift for magnitudes $Ks_{AB} \leq 22$ (resp. 24). 
               All galaxies with $Ks_{AB} \leq 24$ are plotted, above the $Ks-$band completeness magnitude limit (Bielby et al. 2012). }
         \label{sel_bzk}
   \end{figure*}

As the BzK colour-colour selection is commonly used to select galaxies in this
redshift range (e.g. Forster-Schreiber et al. 2009, Daddi et al. 2010, McCracken et al. 2010, Lin et al. 2012), 
it is important to build a robust understanding of the
possible limitations of this technique.
Using the VVDS $i-$band magnitude-selected spectroscopic sample,
we can check the location of galaxies with known redshifts
in a colour-colour plane, and verify which fraction of galaxies 
verifying the BzK selection are indeed in the redshift range $1.4 \leq z \leq 2.5$.
We use the (g-z) versus (z-Ks) distribution of galaxies with reliable 
redshifts in this redshift interval as shown in Figure \ref{sel_bzk}. 
Although our filter set is slightly different, we empirically verify
that the BzK criterion (Daddi et al., 2004) is suitable for gzK, maximizing
the selection of galaxies within this redshift range,
and minimizing the number of galaxies outside (see also Bielby et al., 2012). We have slightly adapted
the criterion to take into account the g-band vs. B band difference.
For 'sgzK' active galaxies analogous to 'sBzK', they verify:
\begin{equation}
(z - Ks )_{AB} - 1.1\times(g - z)_{AB} \geq -0.2
\label{sgzk}
\end{equation}
and for 'pgzK' passive galaxies analogous to 'pBzK' they verify both:
\begin{equation}
(z - Ks )_{AB} - 1.1\times(g - z)_{AB} < -0.2
\label{pgzk1}
\end{equation}
and
\begin{equation}
(z - Ks )_{AB} \geq 2.5
\label{pgzk2}
\end{equation}

% original BzK is $(z − Ks )_{AB} − (B − z)_{AB} \geq -0.2 $

Using the VVDS data set, and only the %430 
galaxies with the most
reliable redshifts (flags 3, 4 and 9), hence independent of any weight correction, 
we find that the gzK criterion is quite
efficient in picking-up galaxies with $1.4 \leq z \leq 2.5$,
as among galaxies with a spectroscopic redshift in this range, 83\% (VVDS Deep) to 94\% 
(VVDS Ultra-Deep) verify criterion (\ref{sgzk})
for $Ks_{AB} \leq 22$. Going to fainter K magnitudes $Ks_{AB}=24$ (above the completeness limit of the 
$Ks-$band imaging, Bielby et al. 2012) our spectroscopic data 
becomes significantly incomplete, leaving only bluer galaxies in
our sample (see Le F\`evre et al. 2013, fig.18), but relying again on the most reliable
redshifts we find that still 78\% (Deep) to 86\% (Ultra-Deep) of galaxies verify criterion (\ref{sgzk})
down to this depth. 
This shows that there is no major break in the BzK/gzK selection ability down to this magnitude, 
but it appears that a significant fraction of the
population of galaxies escape the selection, as also identified from simulations
(Merson et al. 2013). 
%We find that applying this gzK selection blindly from the photometric data
%only would properly select about 90\% of galaxies with $1.4<z<2.5$.
We find that the contamination, computed as the fraction
of galaxies in the selection area  satisfying
criterion (\ref{sgzk}) but at a redshift outside the expected
redshift range, is quite significant at about 
35\% in the Deep survey. This contamination seems to be magnitude
dependent as in the Ultra-Deep survey it is increasing from 23\% to 34\% for limiting 
magnitude from $Ks_{AB}=22$ to $Ks_{AB}=24$. 
%While this might be partly due to  
%photometric errors at these faint magnitudes, 

This analysis mostly concerns the active sBzK galaxies, as, as expected, we find only a few galaxies
verifying the pBzK criterion (\ref{pgzk1})+(\ref{pgzk2}). However, adapting the pBzK criterion to  
$(z - Ks )_{AB} - 1.1\times(g - z)_{AB} < -0.2$
and $(z - Ks )_{AB} \ge 2.25$ would enable access to a small population of galaxies identified with the right
spectroscopic redshift in this category, or about 7\% of our sample. %These galaxies are...

We conclude that the BzK/gzK photometric criterion 
is working well in selecting a sample of star-forming galaxies with $1.4 \leq z \leq 2.5$,
as it selects $\simeq$80\% of galaxies in the correct redshift range.
However, using BzK/gzK photometric selection alone, it would be necessary to take into account that 
a BzK/gzK photometric sample would be significantly contaminated,
with $\sim30$\% of the galaxies selected in the colour-colour selection area
having a redshift outside the $1.4 \leq z \leq 2.5$ range down to $Ks_{AB}=22$,
and a contamination rate possibly increasing at fainter magnitudes.
%One could then decide to adjust the criterion (1) to lower the contamination
%but at the expense of a lower fraction of galaxies within the $1.4 \leq z \leq 2.5$ range.
%For instance, using $(z - Ks )_{AB} - XX\times(g - z)_{AB} \geq XX $ (2)
%would bring down the fraction of galaxies with the right redshift selected by
%the gzK criterion to XX\% but decrease the contamination to XX\%.

\subsection{Galaxies with $2.7 \leq z \leq 3.4$}
\label{z3}

\begin{figure}
   \centering
   \includegraphics[width=8cm]{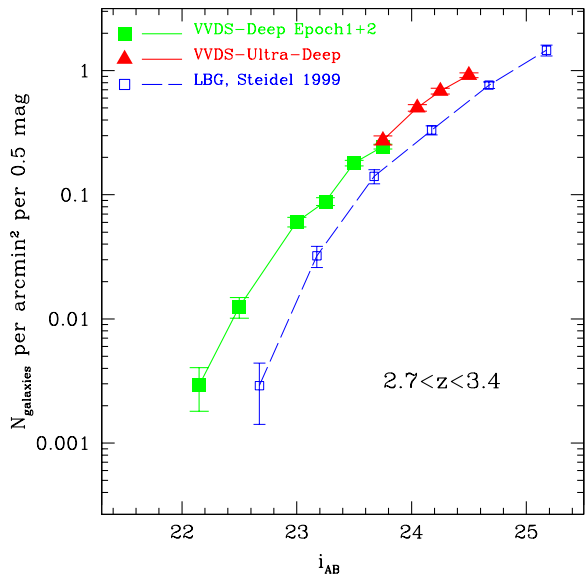}
      \caption{Projected galaxy number counts per square arc-minute as a function of
       $i_{AB}$ magnitude for galaxies with spectroscopic redshifts $2.7 \leq z_{spec} \leq 3.4$, selected solely 
       from their $i_{AB}$ magnitudes, from the 
       VVDS--Deep (filled squares), the VVDS--UltraDeep (triangles), compared to
       data for galaxies selected from the Lyman-break technique (from Steidel et al., 1999; 
       empty squares)}
         \label{dens_z3}
   \end{figure}

The projected number density of galaxies at $2.7 \leq z \leq 3.4$ has been computed
using the full Deep and Ultra-Deep samples. 
%The Deep sample has been corrected for 
%incompleteness using the Ultra-Deep sample, leading to a sample $\simeq97$\% complete.
%The Ultra-Deep sample has been corrected for incompleteness using 
%the method described in Section \ref{sec_zdist} and Le F\`evre et al. (2013), extending the work presented in 
%Le F\`evre et al. (2005b) to a larger and fainter sample. 
Results are shown in Figure \ref{dens_z3}. At a magnitude $i_{AB}=24.5$ we find  $0.94\pm0.10$ galaxy per 
square arc-minute per half magnitude, and a value 15 times lower at $i_{AB}=23.0$, and
in the magnitude range $22 \leq i_{AB} \leq 24.75$ there are $1.72\pm0.15$ gal/arcmin$^2$ (the errors are including 
cosmic variance).
These values are not much sensitive to the SSR described in Section \ref{sel_func}.
Taking into account the estimated uncertainties on the SSR, we expect the true value for 
the projected galaxy counts with $22 \leq i_{AB} \leq 24.75$  to be in the range from 1.4 to 1.85 gal/arcmin$^2$.

There are few spectroscopic surveys in this redshift range to compare to.
In Figure \ref{dens_z3} we compare our results with the LBG counts of Steidel et al. (1999),
using the values given in their Table 3 corrected by the effective volume for a 
cosmology with $\Omega_m=0.3$ and $\Omega_{\Lambda}=0.7$, and transforming
$R_{AB}$ into $i_{AB}$ using the average value of $R_{AB}=i_{AB}+0.09$ derived
from template fitting. The total projected number density from  Steidel
et al. (1999)  in the range $22 \leq i_{AB} \leq 24.75$ is 0.6 gal/arcmin$^2$,
or less  than  half the counts obtained from our magnitude selected sample.
The difference between our counts and the Steidel et al. (1999) counts is
depending on the magnitude of the sample: at bright magnitudes $i_{AB} \leq 23.25$, 
our counts are 3-4 times larger, while at $23.5 \leq i_{AB} \leq 24.75$ the
difference is reduced to a factor $1.5-2$. 
Bielby et al. (2013) performed a spectroscopic survey based on LBG UBR photometric
selection. Down to $R_{Vega}=24.75$, they find about 1 gal/arcmin$^2$, about 40\%
lower than our counts, but we note that their spectroscopic success rate is about 1/3,
making their counts more sensitive to uncertainties on the completeness correction. 
Interestingly, although they are also LBG colour-selected Bielby et al. (2013) identify $1.5-2$
times more  galaxies at relatively bright $R_{Vega} \leq 24.25$ than Steidel et al. (2003), 
but comparable to our results. As the volumes probed by both the Steidel et al. (2003)
and the VVDS are still relatively limited, a possible source of discrepancy, particularly
at the bright end, might be cosmic variance. This will need to be further investigated from
spectroscopic surveys in wider areas.

%{\bf do a comparison with other studies ?}

Barger et al. (2008) claim to disagree with our earlier results (Le F\`evre et al. 2005b),
mis-interpreting the  numbers of galaxies reported in Le F\`evre et al. (2005b)
and Paltani et al. (2007) as the result of a very conservative diagonal limit
in the $u-g,g-r$ colour-colour diagram selection.
However they apparently did not take into account that our sample is magnitude-selected, not 
$ugr$ LBG-selected, so their interpretation of what they believed was  higher numbers than theirs
in this redshift range is incorrect. Moreover, while they claim that they don't find
a population of galaxies outside their $ugr$-LBG selection area, this supposed discrepancy 
with our results is not statistically significant.
Their sample suffers from the small volume sampled corresponding to
the 145 arcmin$^2$ field of view of the  ACS GOODS-North field
and from a significantly brighter completeness magnitude $F850LP_{AB} \sim 23$, more than
1.5 magnitude shallower than ours, which leads to poor number statistics from their sample.
When scaling to a 145 arcmin$^2$ field of view, and a limiting magnitude $F850LP_{AB}=23$, 
we predict from our counts (Figure \ref{dens_z3})
that they should have detected $\simeq10$ galaxies with $2.7 \leq z \leq 3.4$, 
while they find 12, hence entirely compatible with our results.
% given their small number statistics  and large associated uncertainties.

The VVDS counts presented here therefore confirm the trend found by Le F\`evre et al. (2005b) that 
an $i-$band magnitude selected sample identifies a projected number of galaxies 
at $z\sim3$ larger than in a colour-colour LBG sample. 
The Lyman-break technique, or Lyman-$\alpha$ 'drop-out' at the highest redshifts, is at the core of most 
studies of high redshift galaxies beyond redshift 2.5, as the Lyman-break selection is extensively 
used to identify high redshift galaxy samples from the numerous foreground. It is therefore important
to use independent estimates of the efficiency of this selection.
To understand the differences between our magnitude-selected counts and LBG counts, 
we are able to use our $i-$band selected spectroscopic sample 
and identify how many galaxies with known spectroscopic redshifts in the redshift 
range corresponding to
the LBG selection are indeed falling in or out of the LBG colour-colour selection area.
This {\it a posteriori} analysis can then provide an indication of 
the effectiveness and limitations of the photometric LBG colour-colour selection technique.
%The combination of the lyman-limit at 912\AA~ and the IGM absorption
%combine to strongly lower the flux emitted below Lyman-$\alpha$, producing 
%a "break" in the spectrum. When red-shifted into a fixed set of filters, this break
%produces a red colour between two adjacent filters, and is supposed to be  easily
%detected. In a colour-colour plane covering the position of the rest-frame 912--1216\AA~ wavelength 
%domain, galaxies of different spectral types are expected to follow tracks with redshift 
%which all rapidly reach red colours beyond a redshift roughly corresponding to
%the transition wavelength between two adjacent filters. 

\begin{figure}
   \centering
   \includegraphics[width=8cm]{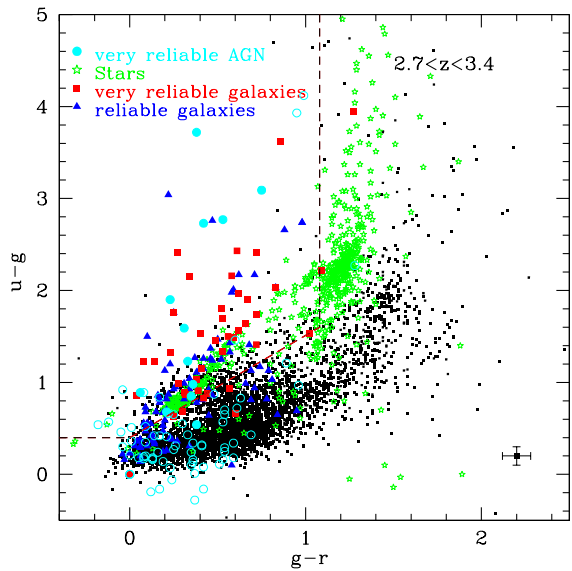}
      \caption{u-g vs. g-r colour-colour diagram for galaxies with spectroscopic
               redshifts $2.7 \leq z \leq 3.4$ magnitude selected in the $i$-band, combining
               the VVDS--Deep and the VVDS-UltraDeep. 
               Galaxies with $2.7 \leq z \leq 3.4$ are identified with triangles and
               squares for reliable flag 2 and very reliable flag 3, 4 and 9, respectively (see text).
               The average colour error at $i_{AB}=24.75$ is indicated on the lower-right corner. 
               46\% of flag 2 and 17\% of the flag 3, 4, 9
               are found outside of the $ugr$ selection area (dashed line) set from template
               tracks. 
               Galaxies with spectroscopic redshifts $z < 2.7$ or $z > 3.4$, are shown
               as dots, they are 3.7 times more numerous than galaxies with spectroscopic 
               redshifts $2.7 \leq z \leq 3.4$ in the $ugr$ selection area. Stars are represented
               with a starred symbol, and also overlap with the LBG selection box.
               Broad-line AGN with flag 13, 14 and 19 are identified as filled circles when $2.7 \leq z_{AGN} \leq 3.4$,
               and by empty circles out of this range. 92\% of the AGN with $2.7 \leq z_{AGN} \leq 3.4$
               are in the $ugr$ selection area, but the contamination is high with about half the AGN in the 
               $ugr$ selection area having a spectroscopic redshift outside $2.7 \leq z \leq 3.4$.
               With this u,g,r filter set, a large contaminating population is therefore expected
               when selecting samples of galaxies with $2.7 \leq z \leq 3.4$  using a priori $ugr$ 
               colour selection. The VVDS magnitude selection is immune to this bias. }
         \label{sel_ugr}
   \end{figure}

As our sample is magnitude selected, it is straightforward to plot the VVDS galaxies
with a spectroscopic redshift $2.7 \leq z \leq 3.4$ on a $u-g,g-r$ colour-colour diagram
and measure the fractions of galaxies in or out of the LBG selection area. 
Very deep $u, g,$ and $r$ photometry is available from the CFHTLS survey, reaching magnitudes 
$u_{AB}=27$, $g_{AB}=26.8$, and $r_{AB}=26.3$ ($\sim5 \sigma$ in 1.2 arc-sec aperture),
significantly deeper than was available in our earlier comparisons (Le F\`evre et al. 2005b).
We present the $u-g,g-r$ distribution of galaxies in the VVDS Deep and Ultra-Deep
samples in Figure \ref{sel_ugr}. We find the following:
\begin{itemize}
\item About 25\% of galaxies with $17.5 \leq i_{AB} \leq 24$ 
and 17\% with $23 \leq i_{AB} \leq 24.75$ and reliable spectroscopic
redshifts in the redshift range $2.7 \leq z \leq 3.4$
fall outside of the $u-g,g-r$ selection area and thus would not be selected from {\it a priori}
$ugr$ selection.
\item For the faintest galaxies in the VVDS Ultra-Deep sample, 73\% of the galaxies which satisfy
the LBG selection criteria in the $u-g,g-r$ diagram have a 
spectroscopic redshift outside the $2.7 \leq z \leq 3.4$ domain, making the 
contamination very high, dominating the galaxies which are
really at $2.7 \leq z \leq 3.4$ by a factor of up to $\sim3.7$ 
(using only our very reliable spectroscopic redshift flags 3, 4 and 9). 
%This 
%contamination increases to about a factor 17 when going to the brighter
%magnitudes of the Deep sample.
\item 92\% of the broad-line AGN with $2.7 \leq z_{AGN} \leq 3.4$
are in the LBG selection area, but the contamination is high with about half the AGN 
with reliable spectroscopic redshift falling in the 
LBG selection area being in reality outside the $2.7 \leq z \leq 3.4$ range.
\item The stellar locus is crossing the LBG selection area, hence a colour-colour
selection using the $u, g, r$ filter set will include a significant fraction of stars: down to
$I_{AB}=24$, $\sim30$\% of the objects satisfying the LBG criterion are stars (the VVDS-0224-04 field 
is at a relatively high Galactic latitude of 58deg, this fraction is likely to change at different latitudes).
%This effect is not present when using the somewhat different bandpasses of the 
%$U_n, G, R$ filter set (see below).
\end{itemize} 
Adjusting the selection boundaries, e.g. increasing the u-g colour cut to
redder colours, would allow to exclude more objects
with  redshifts outside the $2.7 \leq z \leq 3.4$ range, but this would be at the expense of
objects with redshifts in this range. 

\begin{figure}
   \centering
   \includegraphics[width=8cm]{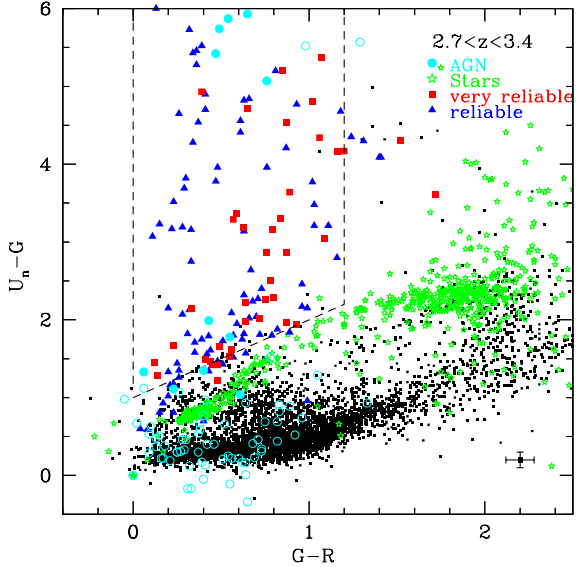}
      \caption{$U_n-G$ vs. $G-R$ colour-colour diagram for galaxies with spectroscopic
               redshifts $2.7 \leq z \leq 3.4$ magnitude selected in the $i$-band, combining
               the VVDS--Deep and the VVDS-UltraDeep. $U_nGR$ magnitudes of galaxies observed
               in the VVDS have been obtained by transforming $ugr$ magnitudes using the
               best galaxy template fitting the full photometric data-set at the measured
               redshift (see text). 
               Galaxies with $2.7 \leq z \leq 3.4$ are identified with triangles and
               squares for confidence classes flag 2 and flag 3, 4 and 9, respectively (see text).
	       The LBG selection of $U_n-G>(G-R)=1$ and $G-R \leq 1.2$ is indicated by the
               dashed line. 
               29\% of flag 2 (reliable redshifts) and 17\% of the flag 3 and 4 (very reliable redshifts)
               are found outside of the $U_nGR$ selection area (dashed line) set from template
               tracks. Among galaxies that would be selected in the $U_nGR$ area as being at $2.7 \leq z \leq 3.4$,
               our spectroscopic redshifts sample shows that 28\% of galaxies are in
               fact at $z < 2.7$ or $z > 3.4$. 
	       Broad-line AGN with flag 13, 14 and 19 are identified as filled circles when $2.7 \leq z_{AGN} \leq 3.4$,
               and by empty circles out of this range. This filter set selects 66\% of the AGN with 
               $2.7 \leq z_{AGN} \leq 3.4$ in the $U_nGR$ selection area, less than the $ugr$ set, but reduces 
               the contamination of AGN in the $U_nGR$ selection area but at other redshifts to $\sim25$\%.
	       Stars are represented
               with a starred symbol, and have a much reduced overlap with the LBG selection box than in the $ugr$ diagram.
               With a $U_nGR$ filter set, we conclude that a sample of galaxies selected
               using a priori $U_nGR$ colour selection, would contain 70\% of galaxies
               with $2.7 \leq z \leq 3.4$, and with 30\% of galaxies and stars forming a contaminating 
               population.  }
         \label{UGR}
   \end{figure}

To test the dependency of the colour selection of objects on the filter 
set, we have transformed our $u, g,$ and $r$ magnitudes into the 
$U_n, G, R$ filter set defined by Steidel \& Hamilton (1993) in the following way. 
We have performed a template fitting of the complete $ugrizJHKs$ data-set we have 
at hands, using the templates to derive the colour corrections to apply to each
$u, g, r$ magnitude in the Gunn system, to transform them to $U_n, G, R$. 
The $U_nGR$ colour-colour diagram is shown in Figure \ref{UGR}. The fraction of
galaxies with redshifts $2.7 \leq z \leq 3.4$ as measured from the VVDS $i-$band magnitude-selected
spectroscopic sample but outside of the
LBG {\it u-g,g-r} selection box is $\sim25$\% in the Deep sample and $\sim17$\% in the Ultra-Deep
sample, for all flags larger or equal to 3, comparable to what we obtain using the $u, g, r$ filters. 
However, the $U_nGR$ filter
set is much more efficient at minimizing the contamination from galaxies
at redshifts outside $2.7 \leq z \leq 3.4$. For galaxies which satisfy the 
$U_nGR$ LBG criteria, 72\% would indeed be
in the right redshift range, while 28\% are galaxies at other redshifts.
This filter set selects 66\% of the broad-line AGN with 
$2.7 \leq z_{AGN} \leq 3.4$ in the $U_nGR$ selection area, less than the $ugr$ set, but reduces 
the contamination of AGN in the LBG selection area which have a 
spectroscopic redshift outside $2.7 \leq z \leq 3.4$ to $\sim25$\%.
The contamination by stars is also low at a few percent. The $U_n, G, R$
filter set is therefore better suited to pre-select galaxies
in the $2.7 \leq z \leq 3.4$ redshift range than the more widely used Gunn  $u, g, r$ set.

%This {\it a posteriori} comparison of the
%colour-colour LBG selection with the magnitude selected spectroscopic sample of the 
%VVDS shows that the effectiveness of the LBG colour-colour
%selection is strongly depending on the exact properties of the filter 
%set  and on the depth of the images used for the photometry,
%and that counts based on colour-colour selection need to be accurately
%corrected from a significant contamination from galaxies at redshifts
%outside the redshift range of interest. 

%To test the impact of the depth of the photometry using colour-colour
%selection, we have followed the locus of galaxies in our magnitude selected 
%spectroscopic sample in the $u, g, r$ colour-colour diagram produced
%using increasingly deeper data from successive CFHTLS releases 
%as new data have been added. We compare the colour-colour
%diagrams using CFHTLS release T003 and release T006 in Figure \ref{cfhtls}.
%We find that...

\begin{figure}
   \centering
   \includegraphics[width=8cm]{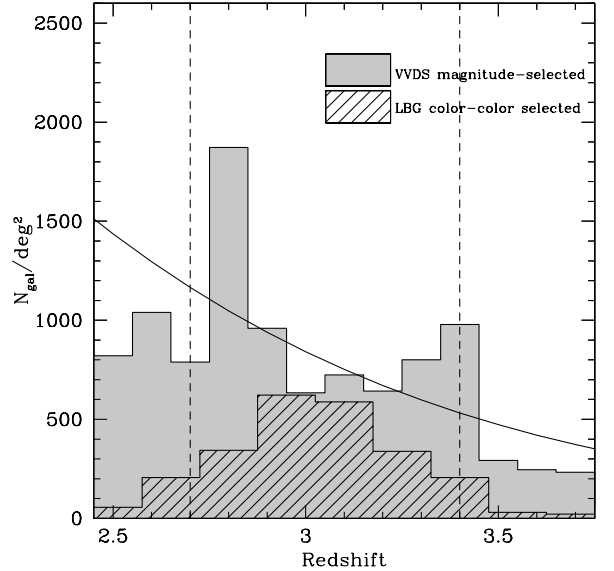}
      \caption{Redshift distribution for galaxies in the range $2.7 \leq z \leq 3.4$, 
       selected solely from their $i_{AB}$ magnitudes from the 
       VVDS--Deep and Ultra-Deep (grey shaded histogram), compared to
       the N(z) for galaxies selected from the Lyman-break technique (from Steidel et al., 1999;
       hatched histogram). The best fit to the VVDS data is represented by the black line.
       The vertical dashed lines represent the limits used to produce and compare counts based
       on the LBG sample. The counts in $2.7 \leq z \leq 3.4$ from the magnitude-selected VVDS are about twice the
       counts from the LBG sample, emphasizing the impact of the colour-colour selection function in
       selecting galaxies from the underlying magnitude-limited population. }
         \label{nz_z3}
   \end{figure}

To understand the higher projected number counts in the VVDS magnitude-selected
sample than the counts from LBG selection, we compare the N(z) from our survey to 
the N(z) of Steidel
et al. (1999) in Figure \ref{nz_z3}, after completeness correction for the 
VVDS and volume correction for the LBG sample. It is clear that while the
magnitude-limited sample shows, as expected, a steady decline of the N(z) with redshift, 
the N(z) of the LBG is bell-shaped, the result of a combination of effects.
The fixed filters band-passes produce a redshift dependent sensitivity 
in identifying a drop in the continuum:
at the middle of the bluer band the sensitivity to identify a continuum drop
is maximum, while when the drop is towards the edges of the bandpass the 
averaged flux in that filter becomes increasingly dominated by either side of the break, lowering the
break contrast. Combined with photometric errors, it results that only galaxies seen 
through increasingly higher IGM absorption are identified when the redshift
places the break increasingly on the sides of the bandpass.
Making a ratio of the two N(z) shows that this  effect 
is largely responsible for the factor of $\simeq2$ difference between the 
magnitude-selected projected counts and the LBG-based counts. 

When counts are transformed into a luminosity function (or mass function,
or any other count-based statistics), taking into
account the  selection function should in principle 
correct any survey for their respective incompleteness to provide 'the same'
LF in the same redshift range. The techniques used to transform counts
into a LF from colour-colour selection are significantly more involved
than assessing the completeness of a spectroscopic sample, as extensive 
photometric simulations need to be performed (e.g. Sawicki \& Thompson
2005, Reddy et al. 2008), which 
include significant uncertainties in terms of the source density (which may lead to
different photometric contamination from neighbours), the morphological properties
(in particular the surface brightness) of the sources, or their spectral 
energy distribution including stars, AGN,  a variable amount of
internal dust or IGM extinction. A detailed computation of 
the luminosity function, luminosity density and star formation rate using the 
counts from the VVDS Deep and Ultra-deep sample is presented
in Cucciati et al. (2012a), including a comparison
of our LF with the LBG LF from Steidel et al. (1999). Based on this 
conservative analysis, we find that the VVDS LF has a slightly lower volume density 
than reported in our earlier findings (Paltani et al., 2007, Tresse et al., 2007), a consequence of 
adding the Ultra-Deep survey to the Deep, and using the deeper u-band photometry that became 
available from the CFHTLS.   However, significant differences between the VVDS-LF
and LBG-LF are still observed, with the magnitude-selected LF from the VVDS being more than 50\% 
higher the the LBG LF at the bright end (Cucciati et al. 2012a, Appendix C).

The comparison of a purely magnitude selected sample with colour-colour selected
samples as presented here shows that the selection of high redshift galaxies
from colour-colour techniques, while powerful to find these galaxies in
significant numbers, is subject to limitations 
and uncertainties which need to be properly estimated and corrected for,
that it is sensitive to the exact filter set used,
and that these errors must be accounted for properly
as they propagate when computing volume complete quantities like 
the global star formation rate density.

%   \begin{figure}
%   \centering
%   \includegraphics[width=8cm]{empty.eps}
%      \caption{Comparison of the locus of magnitude selected VVDS galaxies with $2.7 \leq z \leq 
%3.4$ 
%               in the $u-g, g-r$ colour-colour diagram using the relatively shallow CFHTLS
%               release T003 (left, depth...) and the release T006 (right, depth...). 
%              }
%         \label{cfhtls}
%   \end{figure}

\subsection{Galaxies with $3.4 \leq z \leq 4.5$}

At the highest redshift end of the VVDS, we 
have identified a total of 90 galaxies with reliable redshifts $3.4 \leq z \leq 4.5$ including
73 with flags 2, and 17 with flags [3, 4, 9], and another 110 galaxies with 50\% reliable
flag 1. We find $0.12\pm0.04$ gal/arcmin$^2$ in this redshift range and brighter than $i_{AB}=24$,
and an additional $0.43\pm0.06$ gal/arcmin$^2$ with $24 \leq i_{AB} \leq 24.75$ (errors are including cosmic variance).
Down to $i_{775,AB}=24$ and $i_{775,AB}=24.75$, Bouwens et al. (2007) find $0.025\pm0.01$ and $0.47\pm0.04$ gal/arcmin$^2$,
respectively, in good agreement with our counts, although the
observed colour distribution of our galaxies in $(g-r,r-i)$ shows 
a significant number of galaxies outside the LBG colour-colour
selection area, as described below.
Steidel et al. (1999) report projected counts of about 0.02 and 0.25 gal/arcmin$^2$ down to $I_{AB}=24$
and $I_{AB}=24.75$, respectively, in a redshift interval $3.8 \leq z \leq 4.5$, and after correction 
for their incompleteness, about twice lower than our counts. 

\begin{figure*}
   \centering
   \includegraphics[width=15cm,bb = 18 143 574 457]{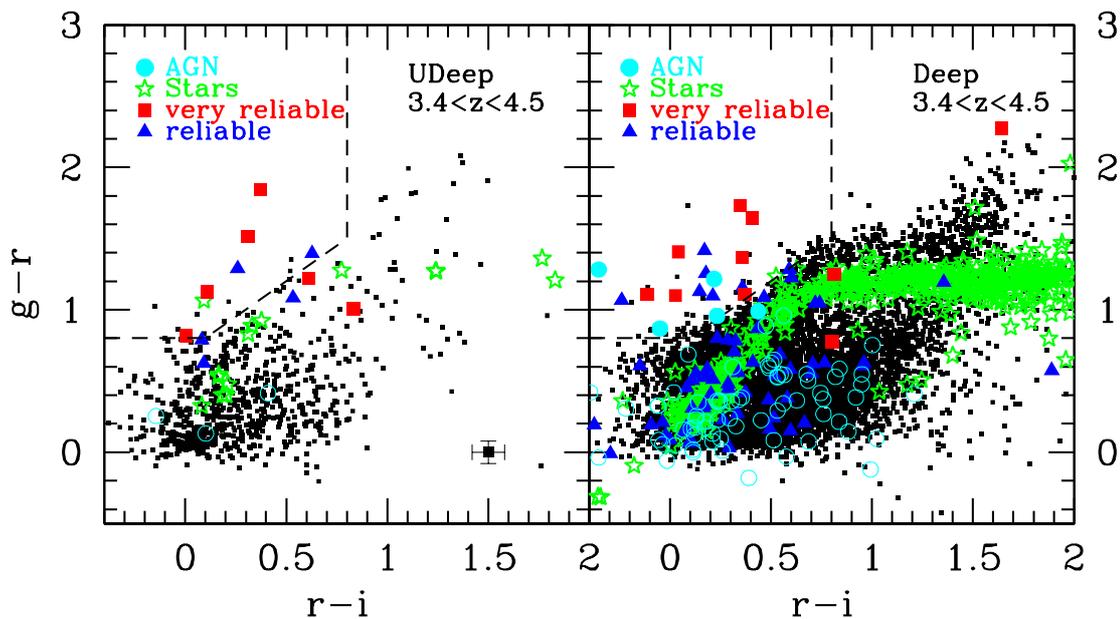}
      \caption{g-r vs. r-i colour-colour diagram for galaxies with spectroscopic
               redshifts $3.4 < z \leq 4.5$ magnitude selected in the $i$-band, for
               the VVDS--Deep (right) and the VVDS-UltraDeep (left). 
               Galaxies with $3.4 < z \leq 4.5$ are identified with triangles and
               squares for confidence classes flag 2 and flag 3, 4 and 9, respectively (see text).
               The typical colour errors at $i_{AB}=24.75$ are identified on the lower-right corner (left panel).
%               80\% of flag 2 (secure redshift) and 83\% of the flag 3 to 9 (very secure redshifts)
%               are found outside of 
               The $gri$ selection area (dashed line) has been set to minimize contamination and maximize
               the number of galaxies in this redshift range. 
               Galaxies with spectroscopic redshifts $z < 3.4$ or $z > 4.5$, are shown
               as dots, they are $\sim6.7$ times more numerous than galaxies with spectroscopic 
               redshifts $3.4 \leq z \leq 4.5$ in the $gri$ selection area. 
	       Broad-line AGN are represented by a circle, filled when the spectroscopic redshift is $3.4 \leq z_{AGN} \leq 4.5$,
               empty when outside this range. 
	       Stars are represented
               with a starred symbol, 80\% of them are excluded from the chosen drop-out galaxy box.
               With this g,r,i filter set, a large contaminating population is therefore expected
               when selecting samples of galaxies with $3.4 < z \leq 4.5$  using a priori $gri$ 
               colour selection.  }
         \label{gri}
   \end{figure*}

The VVDS spectroscopic sample also enables to make
an {\it a posteriori} check of the efficiency of the LBG selection in the $g-r, r-i$ colour-colour plane. 
Based on the distribution of galaxies with redshifts outside  $3.4 \leq z \leq 4.5$ and 
with very reliable redshifts (flags 3, 4, 9),
we have adjusted the LBG colour criterion in the $gri$ filter set to $(g-r) \geq 0.8$, $(r-i) \leq 0.8$
and $(g-r) - (r-i) \geq 0.7$ (Figure \ref{gri}). We find that in this redshift range and with
the $g$, $r$, and $i-$band photometry, it is hard to null the contamination by galaxies
outside of the targeted redshift range without loosing too many of the  
galaxies in the correct redshift range 
$3.4 \leq z \leq 4.5$. With the above colour criteria,  we find from the Ultra-Deep sample that
there are about as many galaxies with redshifts $3.4 \leq z \leq 4.5$ than
galaxies at other redshifts in the LBG box, while  73\% of galaxies selected from
the LBG criterion are outside this redshift range in the Deep sample. We can also estimate the 
fraction of galaxies which have the correct redshift  $3.4 \leq z \leq 4.5$,
but which would not be selected in the LBG box: 
% using again the very reliable redshifts, 
we find that about one third of the galaxies are outside of the
LBG box, both in the Deep and Ultra-Deep samples. The situation seems even worse 
when looking for the reliable redshifts (flag 2), for which 60 to 90\% are outside
of the LBG box (Figure \ref{gri}). 
Only a few broad-line AGN are identified in this redshift range in the Deep survey, 
80\% appear in the $gri$ LBG selection box, and there does not seem to be any contamination
in the $gri$ selection box from AGN at redshifts outside $3.4 \leq z \leq 4.5$.
%CHECK if compatible with SSR.

We conclude from this analysis that using a priori colour selection in the 
$g-r, r-i$ colour-colour plane for galaxies with  $3.4 \leq z \leq 4.5$
would loose about one third of the galaxies in this redshift range outside
of the LBG box, and there would be a contamination from galaxies at other
redshifts ranging from 100\% to 2.8 times larger. 
It is then possible that colour-colour selected samples identify the 
approximate same projected number densities as magnitude selected samples
due to a lucky balance of loosing approximately the same number 
of galaxies out of the colour-colour selection than is gained from galaxies
verifying the selection but being at the wrong redshifts.
Our findings again indicate that 
the purity of a colour-colour selection is strongly dependent on the filter
set used, and must rely heavily on estimates of the completeness from photometric 
simulations and associated uncertainties. 
%{\bf This is compatible or not with LBG spectroscopic surveys in this redshift range.}

%Plot in Figure \ref{gri}

\section{Discussion}
\label{discuss}

From the final VIMOS VLT Deep Survey data release of 10\,765 galaxies with spectroscopic redshifts in
0.61 deg$^2$ on the 0226-04 field (Le F\`evre et al. 2013),
we have computed the redshift distribution N(z) of magnitude-selected samples at different $i-$band depths,
as well as for $J$, $H$ and $Ks-$limited samples. 
We find that for a sample of galaxies with a spectroscopic
redshift measurement and selected down to $i_{AB}=24$,
the redshift distribution peaks at a mean $\bar{z}=0.92$, and 8.2\%
of the galaxies are at $z \geq 2$. Going down to $i_{AB}=24.75$,
the mean redshift of the distribution becomes $\bar{z}=1.15$ and
the high redshift tail becomes more prominent with 17.1\% of
the galaxies at $z\geq2$. Analytic functions  have been fit to
the $i$, $J$, $H$, and $Ks-$ selected samples, providing reference N(z) for future studies. 

Using the N(z) we have produced a complete census of galaxies up to 
$z\simeq4.5$ from projected number counts. The sum of the VVDS
surveys provide a unique opportunity to infer accurate galaxy number counts as the
survey results from a simple magnitude selection, successfully crossing any instrument-induced 
redshift desert. % (from $z\sim1.5$ to $z\sim2.5$ depending on spectrographs setups).
Using the high equivalent completeness of the VVDS sample, we have derived the projected number density of 
galaxies in different redshift ranges: we find $2.07\pm0.12$ gal/arcmin$^2$ at $1.4<z<2.5$ with 
$Ks_{AB}\leq22.5$, $1.72\pm0.15$ gal/arcmin$^2$
in $2.7<z<3.4$ with $i_{AB} \le 24.75$, and $0.59\pm0.09$ gal/arcmin$^2$ in $3.4<z<4.5$
with $i_{AB} \leq 24.75$. Using our magnitude selected spectroscopic sample,
we have been able to test the effectiveness and limitations that an {\it a priori} 
photometric colour-colour selection would lead to.
We find that colour-colour selected samples like BzK or LBG are efficient at selecting galaxies
in the target redshift ranges, but that 20-30\% of galaxies are missed by the colour-colour selection,
depending on the redshift range. Several effects could produce this scattering beyond the
expected colour-colour locus, in particular the contribution from nebular lines which
may significantly affect the observed colours (de Barros et al. 2012). 
The photometric colour-colour selected samples are significantly affected by
contamination from galaxies at other redshifts but scattered inside the colour-colour selection area, the contamination
being strongly dependent on the exact shape of the photometric filters used, on the colour cuts applied,
and on magnitude.
This further emphasizes that the largest source of uncertainty in colour-selected samples is the simulation
of photometric completeness (e.g. Sawicki \& Thompson 2005, Reddy et al. 2008). While standard photometric simulations are built 
from {\it a priori} expectations on the projected number density and on the range of SED distribution  
of distant galaxies, these simulations may have difficulties to properly take into account the
complex morphological properties of galaxies beyond $z\sim1$, and in particular
objects with low surface brightness. This is further complicated by source confusion from 
projected nearby objects, either along the line of sight or physical merging pairs (which 
represent $\sim20$\% of the $z\simeq3$ population, Tasca et al. 2013), or because of
the increasing number of atypical SEDs at $z>1$ due to varying dust attenuation (Cucciati et al. 2012), 
a change in AGN activity, etc..
The comparison with our data shows that the photometric
corrections factors applied to colour-colour selected samples may need to be improved
to match the real numbers of galaxies, as they consistently produce lower galaxy
counts over the redshift range $1.4<z<4.5$. As uncertainties on these counts propagate to the computation of e.g.
luminosity and mass functions, and therefore to the history of star formation and mass assembly
which form the basis of our current understanding of galaxy evolution, more efforts should be
placed on the completeness corrections applied to colour-colour selected samples. On the other hand,
spectroscopic samples obtained from purely magnitude-selection also suffer from incompleteness
depending on the wavelength range and depth, but reaching spectroscopic completeness close 
to 100\% enables an unbiased view of the galaxy population.
Purely magnitude selected samples allow an easier control of the selection function 
and are less dependant on any change in the SED distribution of the population to be observed.
%- colour-colour selection : very efficient to target a specific population for which the SEDs are already known. 

Although the N(z) is a simple statistical description of
the galaxy population, it is combining all evolutionary effects of the different
galaxy populations selected by a survey, and it is therefore 
sensitive to all major physical processes at work in the build-up of galaxies
along cosmic time. The N(z) predicted using dark matter simulations coupled to semi-analytic 
models (De Lucia \& Blaizot, 2007; Wang et al., 2008; Guo et al., 2011, Guo et al., 2013) 
do not accurately represent our observed redshift distribution data. %beyond our measurement uncertainties. 
We find that simulations over-predict the number of low redshift $z<1$ galaxies, by factors of $\sim1.3-1.6$
at $0.7<z<0.9$, $\sim1.2-1.3$ at $0.5<z<0.7$, for galaxies 
brighter than $i_{AB}=24$. 
The discrepancy between models and observations increases with magnitude at low redshifts $z<0.9$,
with 1.8 times more galaxies in the SAM at $0.7<z<0.9$ for $23 \leq i_{AB} \leq 24.75$,
indicating that the simulations overproduce low luminosity objects.  
This points out, as already noted 
from luminosity function analysis (e.g. Cucciati et al. 2012b), that the lowering of the number of 
galaxies from processes like merging or star-formation quenching and feedback is not sufficiently effective in 
these simulations.  
The simulations also significantly under-predict the number of star-forming galaxies at higher redshifts, 
the discrepancy becoming higher with increasing redshift, with $\sim2$ times more galaxies observed 
at $2<z<3$, and $\sim3$ times more at $3<z<4$ down to $i_{AB}=24.75$. This indicates that the formation processes
in simulations are not sufficiently efficient to produce bright galaxies early enough in cosmic time.
The $Ks-$band N(z) comparison to the Henriques et al. (2012) simulation shows a similar behaviour,
but somewhat exacerbated:
the model produces more galaxies at all redshifts $z<2$, with about 40\% more
galaxies in $0.8<z<1$, and not enough galaxies at $z>2$ with about 4 times less
galaxies in the model than observed at $2<z<3$.  These findings on the N(z) for $i-$band 
and $Ks-$band selected samples then indicate that 
the star formation activity in model galaxies is not strong enough to produce
the bright galaxies observed in $i-$band counts, nor, if we take the $K$-band as a proxy for stellar mass,
enough massive galaxies at $z>2$.
While these deficiencies in the models have been already pointed out, the N(z)
provides a simple synthetic view of the main discrepancies between current
simulations and observations, and understanding these discrepancies offers a 
clear opportunity to make progress in our understanding of galaxy evolution.

\section{Summary}
\label{summary}

Using the final VVDS sample we have derived galaxy counts over $0<z<5$. We find that:
\begin{itemize}
\item The redshift distribution N(z) of galaxies magnitude-selected from $i-$band,
or from $J$, $H$ or $Ks-$ bands, can be assembled with a low incompleteness,
successfully crossing any instrument-induced redshift desert.
\item The N(z) of samples with $17.5 \leq I_{AB} \leq 24$, $17.5 \leq i_{AB} \leq 24.75$, and $23 \leq i_{AB} \leq 24.75$,
have mean redshifts $\bar{z}=0.92$, $1.15$, and $1.38$, respectively.
We observe a high redshift tail of bright galaxies up to $z\sim5$, galaxies
with $z \geq 2$ representing $\simeq$17\% of an $i-$band selected sample with $i_{AB} \leq 24.75$.
\item Parametric fit to the N(z) data are provided, offering reference analytical
descriptions of the global galaxy population. These may be used e.g. for weak-lensing analysis requiring a knowledge of the
redshift distribution of background galaxies behind gravitational lenses.
\item We use the N(z) to derive the projected number counts of galaxies
in various redshift ranges. We find $2.07\pm0.12$ galaxies
per arcmin$^2$ at $1.4 \leq z \leq 2.5$ and $Ks_{AB}\leq22.5$, $1.72\pm0.15$ galaxies per
arcmin$^2$ at $2.7 \leq z \leq 3.4$ and $0.59\pm0.09$ galaxies per arcmin$^2$
at $3.4 \leq z \leq 4.5$ brighter than $i_{AB}=24.75$ (errors include Poisson noise as well as cosmic variance).
\item Galaxies identified from magnitude-limited samples are 1.5 to 3 times
more numerous in projected counts than galaxies identified from LBG samples at $z\sim3$,
depending on magnitude. This is mostly because of the varying sensitivity of a filter
set to a continuum break depending on the position of the break in the filter
at the redshift of the source.
\item Combining the VVDS $i-$band selected samples with deep photometry, we 
have analysed the effect of colour-colour selection on galaxy number
counts. We find that BzK/gzK for $1.4 \leq z \leq 2.5$ or LBG selection
using $u-g,g-r$ for $2.7 \leq z \leq 3.4$ and $g-r,r-i$ for  $3.4 \leq z \leq 4.5$,
work well in selecting galaxies in these redshift ranges. However we also identify that
a significant fraction of up to $\sim20-30$\% of galaxies in these redshift ranges
are missed by the colour-colour selection, and, moreover, that the colour-colour
selection also picks-up a large fraction of galaxies at other redshifts
which may seriously contaminate colour-colour selected samples, depending on
the exact shape of the filters used, and the depth of the sample. 
\item Comparing the N(z) with semi-analytic models on the Millennium simulation,
we find that the models overestimate the number of faint galaxies
at low redshifts $z<1$, and under-estimate the number of bright galaxies at high redshifts $z>2$.
\end{itemize}

Our results further stress the importance of building-up robust counts of galaxies
as an input to conducting more evolved analysis of the statistical and physical properties
of galaxies. Accurately counting galaxies remains
a serious challenge at redshifts $z \gtrsim 1$, and associated uncertainties may
have significant impact on our understanding of galaxy formation and evolution. 
Minimizing cosmic variance effects will require observing samples at depth comparable to the VVDS but
over areas much larger than the few square degrees explored so far from spectroscopic surveys.
The improved counts presented in this paper are enabling
a more accurate view of the galaxy population over $0 \leq z \leq 5$,
an important input to prepare for future surveys.

%LOOK AT THE PROPERTIES OF GALAXIES OUTSIDE THE colour BOX: ARE THEY DIFFERENTT
%FROM THE OTHERS (PRESENCE OF CLOSE COMPANIONS, MAGNITUDE, CENTRAL SURFACE BRIGHTNESS, colour, MEAN SPECTRUM...)

%have augmented the VVDS--Deep
%sample of galaxies with a spectroscopic redshift measurement 
%in the range $17.5 \leq i_{ AB} \leq 24$ to XX galaxies, and 
%produced a new sample XX\% complete of XX galaxies 
%with $23 \leq i_{AB} \leq 24.75$. A representative sample 
%of XX galaxies with $17.5 \leq i_{AB} \leq 24$ for which 
%our first epoch observations failed to assign a secure redshift 
%have been re-observed with 4 times more exposure time, enabling 
%a high success rate in redshift measurements and therefore
%enabling to accurately compute the spectroscopic success rate
%of the full VVDS--Deep sample. This sample offers a unique 
%opportunity to conduct an unbiased census of star-forming galaxies 
%with $0\leq z \leq 5$. 

%Kriek et al. 08
%Brammer et al., 2011, ApJ, 739, 24

%Le chinois qui trouve comme nous z=2.

\begin{acknowledgements}
We thank Andrea Cattaneo for useful discussions.
We thank the ESO-Garching and Paranal staff, for following and carrying-out the VIMOS VLT Deep Survey observations.
This work is supported by European Research Council grant ERC-AdG-268107-EARLY. 
The Millennium Simulation databases used in this paper and the web application providing online access to them were constructed as part of the activities of the German Astrophysical Virtual Observatory (GAVO).
 
\end{acknowledgements}

\begin{table*}
\begin{center}
      \caption[]{VVDS counts vs. redshift, in redshift bins of $dz=0.1$, scaled to the number of galaxies per square degree, 
for different magnitude selection}
      \[
 %        \begin{array}{p{0.5\linewidth}lll}
        \begin{array}{lcccc}
           \hline \hline
            \noalign{\smallskip}
            Redshift      &  17.5 \leq I_{AB} \leq 24.0 
                                & 23.0 \leq i_{AB} \leq 24.75 &   
					17.5 \leq i_{AB} \leq 24.75 & 
						Ks_{AB} \leq 22 \\
            \noalign{\smallskip}
            \hline
            \noalign{\smallskip}
0.0  &     188 &      10 &     173 & 0      \\
0.1  &  1\,438 &     148 &  1\,418 & 803    \\
0.2  &  2\,988 &  1\,315 &  4\,048 & 1\,734 \\
0.3  &  5\,579 &  3\,636 &  7\,103 & 3\,110 \\
0.4  &  3\,808 &  2\,388 &  5\,298 & 1\,654 \\
0.5  &  4\,579 &  2\,801 &  6\,915 & 2\,289 \\
0.6  &  7\,207 &  5\,033 & 10\,380 & 3\,675 \\
0.7  &  5\,524 &  3\,956 &  8\,340 & 2\,749 \\
0.8  &  4\,926 &  5\,037 &  8\,867 & 2\,753 \\
0.9  &  7\,552 &  6\,446 & 10\,570 & 3\,933 \\
1.0  &  4\,785 &  7\,313 &  8\,445 & 2\,980 \\
1.1  &  3\,910 &  4\,161 &  6\,488 & 2\,427 \\
1.2  &  3\,119 &  3\,141 &  5\,247 & 2\,333 \\
1.3  &  2\,082 &  3\,244 &  4\,640 & 1\,112 \\
1.4  &  1\,523 &  2\,475 &  2\,616 & 1\,101 \\
1.5  &  1\,556 &  3\,963 &  4\,207 & 684 \\
1.6  &     951 &  1\,510 &  2\,033 & 520 \\
1.7  &     589 &  2\,336 &  1\,743 & 550 \\
1.8  &     512 &  2\,414 &  2\,468 & 220 \\
1.9  &     888 &  1\,574 &  2\,127 & 300 \\
2.0  &     658 &  2\,214 &  2\,020 & 390 \\
2.1  &     752 &  2\,792 &  2\,479 & 325 \\
2.2  &     496 &  2\,285 &  2\,142 & 140 \\
2.3  &     576 &  2\,167 &  2\,248 & 256 \\
2.4  &     448 &  2\,445 &  2\,473 & 175 \\
2.5  &     269 &     715 &     701 & 160  \\
2.6  &     350 &     855 &  1\,003 & 66  \\
2.7  &     540 &     374 &     708 & 143 \\
2.8  &     378 &  1\,468 &  1\,585 & 76  \\
2.9  &     274 &     585 &     798 & 66  \\
3.0  &     119 &     617 &     610 & 14  \\
3.1  &      90 &     566 &     701 & 18  \\
3.2  &     105 &     597 &     645 & 12  \\
3.3  &     120 &     426 &     793 & 43  \\
3.4  &      58 &     862 &  1\,040 & 13  \\
3.5  &     100 &     104 &     261 & 21  \\
3.6  &     100 &      50 &     205 & 37  \\
3.7  &      96 &      20 &     193 & 15   \\
3.8  &      35 &      32 &     152 & 9   \\
3.9  &      28 &      90 &     288 & 9   \\
4.0  &      10 &      22 &     121 & 0   \\
4.1  &      24 &      10 &      16 & 0   \\
4.2  &     2.9 &       0 &      10 & 1   \\
4.3  &     9.4 &       0 &     9.3 & 0   \\
4.4  &      23 &       0 &      20 & 14  \\
4.5  &     2.9 &       0 &      15 & 5   \\
4.6  &     2.7 &       0 &     5.9 & 0   \\
4.7  &     3.1 &       0 &      15 & 1   \\
4.8  &      14 &       0 &     7.5 & 0   \\
4.9  &     2.7 &       0 &      10 & 9   \\
5.0  &       1 &       0 &       5 & 5   \\
            \noalign{\smallskip}
            \hline
         \end{array}
      \]
%\begin{list}{}{}
%\item[$^{\mathrm{a}}$] (\cite{lefevre04})
%\item[$^{\mathrm{b}}$] (\cite{iovino04})
%\item[$^{\mathrm{c}}$] (\cite{radovich})
%\item[$^{\mathrm{d}}$] (\cite{EIS})
%\item[$^{\mathrm{e}}$] (\cite{goods})
%\end{list}
\label{counts}
\end{center}
\end{table*}

\begin{table*}
\begin{center}
      \caption[]{VVDS projected galaxy number counts (galaxies per square arcminute), 
for galaxies with spectroscopic redshifts in selected redshift ranges, as a function of the sample magnitude}
      \[
 %        \begin{array}{p{0.5\linewidth}lll}
        \begin{array}{cccccc}
           \hline \hline
            \noalign{\smallskip}
            i_{AB}      &  \multicolumn{2}{c}{1.4 \leq z \leq 2.5} & \multicolumn{2}{c}{2.7 \leq z \leq 3.4} & 3.5 \leq z \leq 4.5 \\
            magnitude   &  \multicolumn{2}{c}{BzK-like} & \multicolumn{2}{c}{z\sim3 LBG-like} & z\sim4 LBG-like \\
                        &  Deep & Ultra-Deep & Deep & Ultra-Deep & Deep + Ultra-Deep \\
            \noalign{\smallskip}
            \hline
            \noalign{\smallskip}
	18.5-19.0 & 0.002\pm0.001      &                 &               & &  \\
	19.0-19.5 & 0.004\pm0.001      &                 &               & &  \\
	19.5-20.0 & 0.011\pm0.002      &                 &               & &  \\
	20.0-20.5 & 0.026\pm0.004      &                 &               & &  \\
	20.5-21.0 & 0.080\pm0.005      & 0.129\pm0.036   &               & &  \\
	21.0-21.5 & 0.197\pm0.009      & 0.233\pm0.021   &               & &  \\
	21.5-22.0 & 0.307\pm0.010^a    & 0.537\pm0.036   &               & &  \\
	22.0-22.5 & 0.447\pm0.015^a    & 1.165\pm0.047   &               & &  \\
        22.25-22.75 &                  &                 & 0.014\pm0.003 & &  \\
	22.5-23.0 & 0.412\pm0.014^a    & 1.365\pm0.051^a & 0.029\pm0.004 & &  \\
        22.75-23.25 &                  &                 & 0.065\pm0.006 & &  \\
	23.0-23.5 &                    & 1.524\pm0.054^a & 0.094\pm0.007 &                 & 0.032\pm0.004   \\
        23.25-23.75 &                  &                 & 0.190\pm0.010 &                 & 0.043\pm0.005   \\
	23.5-24.0 &                    & 1.246\pm0.049^a & 0.256\pm0.011 & 0.288\pm0.024   & 0.075\pm0.006   \\
	24.0-24.5 &                    &                 &               & 0.703\pm0.037^a & 0.124\pm0.015^a \\
	24.25-24.75 &                  &                 &               & 0.939\pm0.046^a & 0.312\pm0.025^a \\
            \noalign{\smallskip}
            \hline
         \end{array}
      \]
\begin{list}{}{}
\item[$^{\mathrm{a}}$] Counts affected by incompletness (see text)
%\item[$^{\mathrm{b}}$] (\cite{iovino04})
%\item[$^{\mathrm{c}}$] (\cite{radovich})
%\item[$^{\mathrm{d}}$] (\cite{EIS})
%\item[$^{\mathrm{e}}$] (\cite{goods})
\end{list}
\label{counts_zint}
\end{center}
\end{table*}


\begin{thebibliography}{}

   \bibitem[2004]{abraham} Abraham, R., et al., 2004, A.J., 127, 2455   % OK

   \bibitem[2010]{angulo} Angulo R. E., White S. D. M., 2010, \mnras, 405, 143      % OK

   \bibitem[2008]{barger} Barger, A.J., Cowie, L. L., Wang, W.-H., 2008, \apj, 689, 687      % OK

   \bibitem[Behroozi et al. 2013]{behroozi} Behroozi, P.S., Wechler, R., Conroy, C., 2013, \apj, 770, 57  % OK

   \bibitem[2009]{bielby} Bielby, R., et al., 2012, \aap, 545, 23                            % OK

   \bibitem[Bielby et al. 2013]{bielby13} Bielby, R., et al., 2013, \mnras, 430, 425         % OK

   \bibitem[Blain et al. 2002]{blain} Blain, A.W., Smail, I., Ivison, R. J., Kneib, J.-P., \& Frayer, D. T, PhR, 369, 111  % OK

   \bibitem[2008]{blanc08} Blanc, G. A., et al., 2008, \apj, 681, 1099                % OK

   \bibitem[Boone et al. 2011]{boone11} Boone, F., et al., 2011, \aap, 534, 124       % OK

%   \bibitem[2009]{boylan} Boylan-Kolchin M., Springel V., White S. D. M., Jenkins
%   A., Lemson G., 2009, \mnras, 398, 1150

   \bibitem[2007]{bouwens} Bouwens, R.J., Illingworth, G.D., Franx, M., Ford, H., 2007, \apj, 670, 928  % OK

   \bibitem[Bouwens et al. 2010]{bouwens10} Bouwens, R.J. 2010, \apj, 709, 133          % OK

   \bibitem[Capak et al. 2011]{capak11} Capak, P., et al., 2011, \apj, 730, 68          % OK

   \bibitem[2011]{cassata11} Cassata, P., Le F\`evre, O., et al., 2011, \aap, 525, 143  % OK

   \bibitem[Cimatti et al. 2002]{cimatti02} Cimatti, A., et al.,  2002, \aap, 381, 68   % OK

   \bibitem[2008]{cimatti} Cimatti, A., et al., 2008, \aap, 482, 21                    % OK

   \bibitem[Contini et al. (2012)]{contini} Contini, T., et al., 2012, \aap, 539, 91   % OK

   \bibitem[Coupon et al. 2009]{coupon09} Coupon, J., et al., 2009, \aap, 500, 981   % OK

   \bibitem[2012]{cucciati12a} Cucciati, O., et al., 2012a, \aap, 539, 31            % OK

   \bibitem[2012]{cucciati12b} Cucciati et al 2012b, \aap, 548, 108                 % OK

   \bibitem[2012]{cuillandre} Cuillandre, J.C., et al., 2012, Proc. SPIE 8448, Observatory Operations: Strategies, Processes, and Systems IV, 84480  % OK

   \bibitem[2004]{daddi} Daddi, E., et al., 2004, \apj, 617, 746                    % OK

   \bibitem[2010]{daddi10} Daddi, E., et al.,  2010, \apj, 713, 686                 % OK

   \bibitem[2012]{debarros12} de Barros, S., Schaerer, D., Stark, D. P., 2012, arXiv:1207.3663  % OK

   \bibitem[2011]{sdlt} de la Torre, S., et al., 2011, \aap, 525, 125               % OK

   \bibitem[2007]{delucia} De Lucia, G. \& Blaizot, J. 2007, MNRAS, 375, 2          % OK

   \bibitem[Ellis et al. 2013]{ellis13} Ellis, R.S., et al.,  2013, \apj, 763, 7    % OK

   \bibitem[2007]{faber} Faber, S.M., et al., 2007, \apj, 665, 265                  % OK

   \bibitem[Forster-Schreiber et al. 2009]{fs09} Forster-Schreiber, N., et al., 2009, \apj, 706, 1364   % OK

   \bibitem[2003]{franx} Franx, M., et al., 2003, \apj, 587, L79                       % OK

   \bibitem[2008]{fu} Fu, L., et al., 2008, \aap, 479, 9                               % OK

   \bibitem[Garilli et al. (2008)]{garilli} Garilli, B., et al., 2008, \aap, 486, 683  % OK

   \bibitem[Guo et al. (2011)]{guo} Guo, Q., White, S., Boylan-Kolchin, M., 
   De Lucia, G., Kauffmann, G., Lemson, G., Li, C., Springel, V., Weinmann, S.,
   2011, MNRAS, 413, 101                                                              % OK

   \bibitem[2012]{guo12} Guo, Q., et al., 2013, \mnras, 428, 1351                      % OK

   \bibitem[Guzzo et al. (2013)]{guzzo13} Guzzo, L., et al., 2013, arXiv:1303.2623     % OK

   \bibitem[Hildebrandt et al., 2012]{Hidebrand13} Hilbebrandt, H., et al., 2012, \mnras, 421, 2355   % OK

   \bibitem[1998]{hu} Hu, E.M.; Cowie, L.L.; McMahon, R.G., 1998, \apj, 502, 99          % OK

   \bibitem[Ilbert et al 2004]{ilbert04} Ilbert, O., et al.,  2004, \mnras, 351, 541     % OK

   \bibitem[2005]{ilbert05} Ilbert, O., et al., 2005, \aap, 439, 863                     % OK

   \bibitem[Ilbert et al. (2006)]{ilbert06} Ilbert, O., et al., 2006, \aap, 457, 841     % OK

   \bibitem[Ilbert et al. (2009)]{ilbert09} Ilbert, O., et al., 2009, \apj, 690, 1236    % OK

   \bibitem[Ilbert et al. (2010)]{ilbert10} Ilbert, O., et al.,  2010, \apj, 709, 644    % OK

   \bibitem[Ilbert et al. (2013)]{ilbert13} Ilbert, O., et al., 2013, \aap, in press (arXiv:1301.3157)  % OK
  
   \bibitem[2011]{komatsu} Komatsu E., Smith K. M., Dunkley J., Bennett C. L., 
   Gold B., Hinshaw G., Jarosik N., Larson D., et al. 2011, \apjs, 192, 18               % OK

   \bibitem[Kong et al. (2006)]{kong06} Kong, X., et al. 2006, \apj, 638, 72             % OK

   \bibitem[1995]{olf95} Le F\`evre, O., et al., 1995, \apj, 475, 60                       % OK

   \bibitem[Le F\`evre et al. 2003]{olf03} Le F\`evre, O., et al., 2003, SPIE, 4841, 1670   % OK

   \bibitem[Le F\`evre et al. (2004)]{olfima} Le F\`evre, O., et al., 2004, \aap, 417, 839  % OK

   \bibitem[Le F\`evre et al. (2005a)]{olf2} Le F\`evre, O., et al., 2005a, \aap, 200, 58  % OK

   \bibitem[Le F\`evre et al. (2005b)]{olf1} Le F\`evre, O., et al., 2005b, Nature, 437, 519 % OK

   \bibitem[Le F\`evre et al. (2013)]{olf3} Le F\`evre, O., et al., 2013, arXiv:1307.0545  % OK

   \bibitem[Lemson]{Lemson06} Lemson, G., \& the Virgo Consortium 2006, arXiv:astro-ph/0608019  % OK
 
   \bibitem[Lilly et al. (1995)]{lilly1} Lilly, S.J., Le F\`evre, O., Crampton,
      D., Hammer, F., \apj, 1995, 455, 50                                                  % OK

   \bibitem[Lilly et al. (2007)]{lilly2} Lilly, S.J., Le F\`evre, O., et al., 2007, \apjs, 172, 70  % OK

   \bibitem[Lin et al. 2012]{lin12} Lin, L., et al., 2012, \apj, 756,71                  % OK

   \bibitem[McCracken et al. 2003]{hjmcc} McCracken, H. J., Radovich, M., Bertin, E., Mellier, Y., 
    Dantel-Fort, M., Le F\`evre, O., Cuillandre, J. C., Gwyn, S., Foucaud, S., Zamorani, G., 2003, \aap, 410,17   % OK

   \bibitem[2009]{hjmcc09} McCracken, H.J., et al., 2009, \apj, 708, 202                   % OK

   \bibitem[2010]{hjmcc10} McCracken, H.J., et al., 2010, \apj, 708, 202                 % OK

   \bibitem[2012]{hjmcc12} McCracken, H.J., et al., 2012, \aap, 544, 156                 % OK

   \bibitem[McLure et al. 2010]{mclure} McLure, R.J., et al.,  2010, \mnras, 403, 960    % OK

   \bibitem[Merson et al. 2013]{merson13} Merson, A.I., et al., 2013, \mnras, 429, 556   % OK

   \bibitem[Moster et al. 2011]{moster11} Moster, B.P., Somerville, R.S., Newman, J.A., Rix, H-W., 2011, \apj, 731, 113   % OK

   \bibitem[Oliver et al. 2010]{oliver} Oliver, S., et al., 2010, \aap, 518, 21         % OK

   \bibitem[2008]{ouchi} Ouchi, M., et al.,  2008, \apjs, 176, 301                      % OK

   \bibitem[Pirzkal et al. 2013]{pirzkal13} Pirzkal, N., et al., 2013, arXiv:1304.4594   % OK

   \bibitem[Planck 2013]{planck13} Planck collaboration,  2013, arXiv1303.5076          % OK

   \bibitem[2005]{reddy2} Reddy, N., et al., 2005, \apj, 633, 748                       % OK

   \bibitem[2008]{reddy} Reddy, N., et al., 2008, \apjs, 175, 48                        % OK

   \bibitem[2005]{sawicki} Sawicki, M., and Thompson, D., 2005, \apj, 635, 100     % OK

   \bibitem[2005]{schi} Schiminovich, et al., 2005, \apj, 619, 47                  % OK

   \bibitem[2010]{schra} Schrabback, T., et al., 2010, \aap, 516, 63               % OK
 
   \bibitem[2004]{sommer} Sommerville, R., et al., 2004, \apj, 600, 171            % OK

   \bibitem[2003]{spergel} Spergel D. N., Verde L., Peiris H. V., et al., 2003, \apjs, 148, 175       % OK

   \bibitem[2005]{springel} Springel V., White S. D. M., Jenkins A., et al., 2005, Nature, 435, 629   % OK

   \bibitem[1999]{steidel93} Steidel, C.C., \& Hamilton, D., 1993, \aj, 105, 2017                     % OK

   \bibitem[Steidel et al. 1996]{steidel96} Steidel, C.C., Giavalisco, M., Pettini, M., Dickinson, M., Adelberger, K.L., 1996, \apj, 462, 17    % OK

   \bibitem[2003]{steidel03} Steidel, C.C., et al., 2003, \apj, 592, 728        % OK 

   \bibitem[1999]{steidel99} Steidel, C.C., et al., 1999, \apj, 519, 1          % OK

   \bibitem[2007]{tresse} Tresse, L., et al., 2007, \aap, 472, 403              % OK

   \bibitem[2006]{vandokkum} van Dokkum, P.G., et al.,  2006, \apj, 638, 59     % OK

   \bibitem[2001]{waerbeke} van Waerbeke, L., et al., 2001, \aap, 374, 757      % OK

   \bibitem[2008]{wang} Wang, J., De Lucia, G., Kitzbichler, M., White, Simon D.M., 2008, \mnras, 384, 1301   % OK

   \bibitem[2008]{wilkins} Wilkins, S., et al., 2008, MNRAS, 385, 687           % OK

   \bibitem[2001]{wilson} Wilson, G., Kaiser, N., Luppino, G.A., Cowie, L.L., 2001, \apj, 555, 572   % OK

   \bibitem[Wolf et al., 2003]{wolf03} Wolf, C., Meisenheimer, K., 
   Rix, H.-W., Borch, A., Dye, S., Kleinheinrich, M., 2003, \aap, 401, 73   % OK

\end{thebibliography}
\end{document}